\documentclass[12pt]{article}
\pdfoutput=1
\usepackage{jcapmod}

\usepackage[margin=2cm]{geometry}

\usepackage{booktabs}
\usepackage[english]{babel}
\usepackage{amsmath,amssymb,amsbsy,amstext, amsthm, simplewick, amsfonts}
\usepackage{hyperref}
\usepackage{graphicx}
\usepackage[small]{caption}
\usepackage{siunitx}
\usepackage{upgreek}
\usepackage{framed}
\usepackage{wrapfig}
\usepackage{slashed}
\usepackage{afterpage}

\usepackage{multirow}
\usepackage{bbm}
\usepackage[svgnames,dvipsnames,x11names]{xcolor}
\usepackage[utf8x]{inputenc}
\usepackage{selinput}
\usepackage{bm}
\usepackage{float}
\usepackage{mathtools}
\usepackage{subfiles}
\usepackage{multicol}
\usepackage{subcaption}

\usepackage[framemethod=default]{mdframed}
\newmdenv[skipabove=7pt,
skipbelow=7pt,
rightline=false,
leftline=false,
topline=false,
bottomline=false,
backgroundcolor=gray!10,
linecolor=gray,
innerleftmargin=5pt,
innerrightmargin=5pt,
innertopmargin=5pt,
innerbottommargin=5pt,
leftmargin=0cm,
rightmargin=0cm,
linewidth=4pt]{eBox}

\usepackage{colortbl}
\definecolor{lightgreen}{cmyk}{0.2, 0, 0.2, 0.2}
\definecolor{lightgray}{cmyk}{0.1,0.2,0,0.1}
\definecolor{lightgray2}{cmyk}{0.1,0.1,0,0.1}

\newcommand{\minus}{{\scalebox {0.75}[1.0]{$-$}}}

\makeatletter
\newlength{\apb@width}
\newcommand{\autoparbox}[2][c]{\settowidth{\apb@width}{#2}\parbox[#1]{\apb@width}{#2}}

\makeatother

\newcommand{\ud}{\mathrm{d}}
\newcommand{\lab}[1]{{\mathrm{#1}}}
\newcommand{\mb}[1]{{\mathbf{#1}}}
\newcommand{\hodge}{\mathord{*}}

\newcommand{\sups}[1]{{\scriptscriptstyle #1}}
\newcommand{\sminus}{\scalebox {0.65}[1.0]{$\scriptstyle-$}}

\makeatletter
\newsavebox\myboxA
\newsavebox\myboxB
\newlength\mylenA

\newcommand*\xoverline[2][0.75]{%
    \sbox{\myboxA}{$\m@th#2$}%
    \setbox\myboxB\null% Phantom box
    \ht\myboxB=\ht\myboxA%
    \dp\myboxB=\dp\myboxA%
    \wd\myboxB=#1\wd\myboxA% Scale phantom
    \sbox\myboxB{$\m@th\overline{\copy\myboxB}$}%  Overlined phantom
    \setlength\mylenA{\the\wd\myboxA}%   calc width diff
    \addtolength\mylenA{-\the\wd\myboxB}%
    \ifdim\wd\myboxB<\wd\myboxA%
       \rlap{\hskip 0.5\mylenA\usebox\myboxB}{\usebox\myboxA}%
    \else
        \hskip -0.5\mylenA\rlap{\usebox\myboxA}{\hskip 0.5\mylenA\usebox\myboxB}%
    \fi}
\makeatother

\def\beq{\begin{equation}}
\def\eeq{\end{equation}}

\allowdisplaybreaks[1]
\begin{document}
%TITLE PAGE=============================

\begin{titlepage}
\setcounter{page}{1} \baselineskip=15.5pt \thispagestyle{empty}
\bigskip\

\vspace{40pt}

\begin{center}
{\fontsize{20}{18} \bf Real-Time Corrections to the Effective Potential}\\[14pt]
\end{center}

\vskip 50pt
\begin{center}
{Guilherme L.~Pimentel and John Stout}

  \vskip8pt
 %$^1$ 
 Institute of Theoretical Physics, University of Amsterdam,\\Science Park 904, Amsterdam, 1098 XH, The Netherlands
\end{center}

\vspace{1.2cm}
\begin{center} { \bf Abstract} \end{center} 
\vspace{-0.2cm}
\begin{abstract}
\noindent
\end{abstract}

 Non-perturbatively generated effective potentials play an extremely useful and often critical role in string and inflationary model building. These potentials are typically computed by methods that assume the system is in equilibrium. For systems out of equilibrium, like an inflaton rolling down its potential, there are corrections to the semi-classical evolution due to transient phenomena. 
 We provide a new qualitative and quantitative understanding of non-perturbative effects in real time for a wide class of toy quantum mechanical models. We derive an effective Schr\"{o}dinger equation that does not rely on any notion of equilibrium and captures the low-energy dynamics supposedly described by the effective potential. We find that there are potentially large corrections to this potential that are not captured by standard equilibrium techniques, and quantify when these corrections significantly alter the effective dynamics.

\vskip 10pt
%\hrule
\vskip 10pt

\vspace{1cm}
\noindent

\noindent
\noindent

\end{titlepage}

\newpage
%\hrule
%\vspace{0.2cm}
\tableofcontents
%\vspace{0.5cm}
%\hrule

\newpage

%=======================================
\newpage

\section{Introduction}
%=======================================

We are very fortunate our universe is in a state of almost equilibrium. Not only is it beneficial for the formation of intelligent life, but physical processes are much easier to understand if they are a small departure from a steady state. Physicists have had great success in leveraging both to construct accurate low-energy effective field theories that describe a wide range of physical phenomena.  The power of these effective descriptions is in that heavy, unknown degrees of freedom---under the assumption that they eventually return to their vacuum---can be decoupled, with their effect incorporated into an effective action for the low energy theory. However, this assumption may be violated when a system is driven out of equilibrium. For example, spontaneous particle production can occur in cosmological backgrounds, and the notion of integrating out heavy degrees of freedom is at best approximate when they can appear at late times. Time-dependent phenomena can thus modify an effective description.  

In this paper, we study these corrections in a particular class of models whose semi-classical dynamics are wholly determined by a non-perturbatively generated effective potential. These potentials are an often critical tool in the construction of both UV-complete inflationary models and realistic string compactifications. Though they are computed assuming the system is in equilibrium, they are often used to study out-of-equilibrium processes---like an inflaton or quintessence field rolling down its potential---and may receive potentially disastrous time-dependent corrections. Given their role in modern inflationary and string phenomenology, it is thus crucial to understand how these potentials change when the system is allowed to evolve. 
Our ultimate goal, then, is to understand how non-perturbative physics affects real-time dynamics, and how to incorporate transient phenomena in such effective descriptions.

Our motivation for studying these time-dependent corrections comes primarily from cosmology, and in particular natural inflation \cite{Freese:1990rb, Adams:1992bn}. Here, a (pseudo)scalar axion $\phi$ is coupled to other degrees of freedom such that, classically, it enjoys a continuous shift symmetry $\phi \to \phi + \epsilon$. However, this shift symmetry is ``broken'' to a discrete subgroup $\phi \to \phi + 2 \pi f$ by either the formation of a condensate or nonperturbative effects. The low-energy effective theory is typically written as
\begin{equation}
	\mathcal{L} = \minus \frac{1}{2}(\partial \phi)^2 + \Lambda^4 \left(1 - \cos \phi/f\right) + \dots \label{eq:naturalInflation}
\end{equation}
where $\Lambda$ is a dynamically generated scale and $f$ is called the axion decay constant. When coupled to gravity,\footnote{In this paper, we ignore gravity and do not incorporate effects like Hubble friction or a dynamical scale factor. Nonetheless, studying the evolution of a semiclassical ``rolling'' trajectory, sans gravity, will help us understand how the system behaves when spacetime is allowed to evolve.} the axion drives inflationary expansion by slowly rolling down its dynamically-induced effective potential. Since the inflationary scenario is inherently out of equilibrium, transient phenomena will correct the axion's trajectory and may be violent enough to halt inflation entirely.

 Though this effective theory has many UV realizations,\footnote{For instance, see \cite{Adams:1992bn, ArkaniHamed:2003mz, ArkaniHamed:2003wu, Banks:2003sx, BlancoPillado:2004ns, Kim:2004rp, Grimm:2007hs}.} we are solely interested in completions where the effective potential is generated by non-perturbative effects. Throughout this paper we will study quantum mechanical analogues of a specific prototypical UV completion of (\ref{eq:naturalInflation}),
	\begin{equation}
		S = \int\!\ud^4 x\left[ \minus \frac{1}{2} (\partial \phi)^2 - \frac{1}{g^2} \,\lab{tr}\left(F \wedge \hodge F\right) + \frac{\phi}{8\pi^2 f}  \,\lab{tr}\left( F\wedge F\right)  + \mathcal{L}_\lab{cm}\right]. \label{eq:prototype}
	\end{equation}
	where $F$ is the field strength of a compact non-Abelian gauge field $A$ which couples to charged matter described by $\mathcal{L}_\lab{cm}$. Under an infinitesimal shift $\phi \to \phi+ \epsilon f$, this action famously transforms up to a total derivative,
	\begin{equation}
		\Delta S = \frac{\epsilon}{8 \pi^2} \int\lab{tr}\left(F \wedge F\right) = \epsilon \int \lab{tr}\left(\ud \hodge J_\lab{CS}\right),
	\end{equation}
	where $J_\lab{CS}$ is the Chern-Simons current. This term is only sensitive to the topology of the gauge field, which is characterized by a winding number $\int\lab{tr}\left(\ud \hodge J_\lab{CS}\right) \in \mathbbm{Z}$ for the simplest non-trivial topologies. The classical equations of motion are thus invariant under such a shift, so the axion may sit still anywhere along its field space.

	The quantum story is different. Because the winding number is an integer, the Feynman measure $\exp(i S)$ is only invariant under shifts generated by $\phi \to \phi + 2 \pi f$. This matters for the quantum theory, since the gauge field's compactness forces us to sum over all topologically non-trivial field configurations, each of which is weighted differently under an infinitesimal shift $\phi \to \phi + \epsilon f$. This sum over topological sectors thus breaks the continuous shift symmetry down to a discrete subgroup, inducing an effective potential for the axion that can be identified with the gauge theory's vacuum energy. Typically, this is computed by evolving for an infinite amount of Euclidean time, and reading off the large-time asymptotics of the partition function
	\begin{equation}
		V_\lab{eff}(\phi) = \lim_{\beta \to \infty} -\frac{1}{\beta \mathcal{V}}\,\log\, \lab{tr}_A \, e^{-\beta \mathcal{H}(\phi)}\,, \label{eq:veffIntro}
	\end{equation}
	where $\mathcal{V}$ is the spatial volume, $\mathcal{H}(\phi)$ is the gauge theory Hamiltonian as a function of constant $\phi$, and $\lab{tr}_A$ is a sum over all gauge field configurations.
	
	 \begin{figure}
		 	\centering
		 		\includegraphics[scale=1, trim=0 0 0 0]{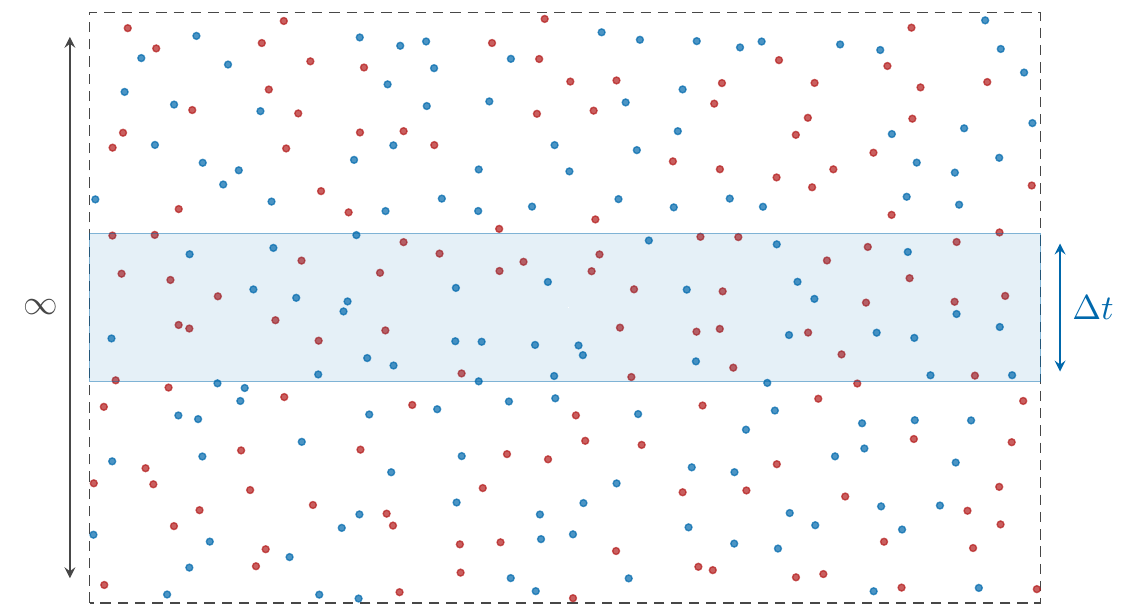}
		 	\caption{A schematic representation of the dilute instanton gas approximation. The effective potential is a collective effort of all types of instanton configurations at all times. We expect that this collective effort is diminished at short times, as highly wound, widely-separated configurations cannot ``fit'' into the time interval $\Delta t$.
		 	\label{fig:DIGA}}
		 \end{figure}
		 
	The primary motivation for this work comes from the following thought experiment. The effective potential receives contributions from gauge field configurations of all winding numbers, and is typically computed using the \emph{dilute instanton gas approximation} \cite{Rajaraman:1982is, Coleman:1985rnk, Polyakov:1987ez, Shifman:2012zz}. This assumes that (\ref{eq:veffIntro}) is dominated by combinations of widely separated, singly-wound (anti)instantons that are well-localized in space and time.
	To approximate $V_\lab{eff}(\phi)$, we then consider a \emph{gas} of these instantons, integrating over all allowed times and positions,\footnote{We  may also need to integrate over other instanton (quasi)zero modes, like the size of the instanton and its orientation in the gauge group.} as schematically illustrated in {Figure~\ref{fig:DIGA}}. Importantly, the effective potential is a collective effort of \emph{all} instantons at \emph{all} times. What happens at short times? Naively, we might think that there are fewer instantons that contribute to the path integral or that contributions of high winding number change behavior, as it is no longer valid to assume that they can be decomposed into widely separated singly-wound events. So, the structure of the effective potential---and thus the dynamics inferred from it---could change dramatically if the axion is moving quickly.
	
	Indeed, if we think of the axion's evolution as a sequence of steps between equilibria, we might expect that it must wait long enough for the instantons to ``fill in'' the effective potential before moving onto its next step.  Since these instantons can be thought of as tunneling events---whose rate is non-perturbatively suppressed in the gauge coupling---we might guess that the axion must move \emph{very} slowly for it to follow the effective potential. We will confirm this intuition in \S\ref{sec:fail}, though we will find other effects that are not obviously captured by this intuitive argument.
	
	 We focus on UV completions like (\ref{eq:prototype}) in particular because they most directly make contact with string theoretic realizations of natural inflation.\footnote{These systems are interesting on a conceptual level, even without the connection to inflation, as theories whose semi-classical dynamics are determined completely by quantum mechanical effects. From this point of view, we expect our findings could also be useful for condensed matter systems driven out of equilibrium.} Typically the axion $\phi$ descends from the dimensional reduction of a $p$-form gauge field along a topologically non-trivial cycle 
	  and receives its potential from either gauge theory instantons or Euclidean D-branes.\footnote{For a review of inflationary models in string theory, see \cite{Baumann:2014nda}.} The redundancy $\phi \sim \phi + 2 \pi f$ is then a descendant of the higher-dimensional gauge symmmetry, which protects the axion's potential from plausibly dangerous quantum gravitational corrections. These models thus represent an important, well-controlled lamppost for realizing large-field inflation in string theory. It is not clear whether quantum gravity allows for non-violent super-Planckian field displacements and these effective potentials have been at the heart of an on-going debate \cite{Banks:2003sx, Svrcek:2006yi, Rudelius:2014wla, Bachlechner:2014gfa, Montero:2015ofa, Bachlechner:2015qja, Brown:2015iha, Brown:2015lia, Rudelius:2015xta, Junghans:2015hba, Heidenreich:2015wga, Conlon:2016aea, Baume:2016psm,  Hebecker:2017uix, Palti:2019pca}. It would be useful, then, to understand how these potentials behave in the regimes where they are actually used.\footnote{A non-equilibrium understanding of non-perturbatively generated effective potentials would also help elucidate their role in string compactifications. The validity of the approximations used in establishing that non-perturbative effects controllably stabilize moduli
	   has been the subject of recurring debate (see, e.g.  \cite{Sethi:2017phn, Kachru:2018aqn}). Though our findings are only tangentially related, we believe that developing an out-of-equilibrium picture of these effects will provide a better understanding of when they are under control.}

	\subsubsection*{Outline}
		 In \S\ref{sec:ToyModels}, we describe a class of $(0+1)$-dimensional toy models with an ``axion'' $\varphi$ and ``gauge field'' $A$ that, we argue, capture the relevant features of the prototype (\ref{eq:prototype}).\footnote{We are primarily interested in the dynamics of the \emph{zero modes} of quantum fields in instanton-induced effective potentials. We will not be concerned with the theory of fluctuations around such trajectories. This focus on zero modes makes the connection between these toy models and their more realistic four-dimensional brethren less far-fetched.} 
	We show that topologically non-trivial gauge field configurations generate an effective potential for the axion and that this class smoothly interpolates between potentials that are either ``monodromy-like'' or ``instanton-like.'' Monodromy-like potentials (c.f. Figure \ref{fig:monodromyPotential}) are roughly a sum over branches, schematically
		\begin{equation}
			V_\lab{eff}(\varphi) \sim \min_{\ell \in \mathbbm{Z}} V(\varphi - \ell)\,,
		\end{equation}
		while instanton-like potentials are well-approximated by a single cosine (c.f. Figure \ref{fig:mathieuEnergies})
		\begin{equation}
			V_\lab{eff}(\varphi) \sim \Lambda e^{\sminus \mathcal{S}_1} (1 - \cos 2 \pi \varphi) + \mathcal{O}(e^{-2 \mathcal{S}_1})\,,
		\end{equation} 
		where $\mathcal{S}_1$ is the action of a single instanton.
		We illustrate our results in the monodromy-like \emph{gapless model} and an instanton-like \emph{gapped model}.  Throughout, we pay special attention to the boundary conditions imposed in deriving these effective potentials, and argue that the process of integrating out the gauge field assumes boundary conditions that are not realized when the axion is allowed to evolve.

		 In \S\ref{sec:effSchrodinger}, we derive an effective Schr\"odinger equation that describes the exact dynamics of the system. This presentation makes it clear which degrees of freedom should be discarded to derive a low-energy effective description. Schematically, it is given by
		 \begin{align}
				i \hbar\, \partial_t \Phi_{0}(\varphi,t) &= \left(\frac{p_\varphi^2}{2 f^2}   + V_\lab{eff}(\varphi) + V_{0,0}(\varphi)\right) \Phi_{0}(\varphi,t)  + \sum_{n' \neq 0} \left[F_{0, n'}(\varphi) \, p_\varphi  + V_{0,n'}(\varphi) \right]\Phi_{n'}(\varphi, t)\,,
			\end{align}
		where $\Phi_0$ and $\Phi_{n' \neq 0}$ are the axion $\varphi$'s wavefunction when the gauge field $A$ is in its vacuum and higher excited states, respectively. At low energies, the axion evolves according to a potential induced by the gauge field. This potential is not just the effective potential $V_\lab{eff}(\varphi)$ derived by equilibrium methods, but is corrected by a new term $V_{0, 0}(\varphi)$ that only disappears when the axion is non-dynamical. At higher energies, the axion can excite the gauge field out of its vacuum and this is encoded in the $\varphi$-dependent ``friction''-like couplings $F_{0, n'}$ and $V_{0, n'}$. This effective Schr\"{o}dinger equation allows us to derive speed limits for the axion, above which the description breaks down.
		
		We will find that this effective description is highly dependent on our choice of initial state. How do we know we have the right one? Furthermore, this description does nothing to elucidate how the gauge field instantons behave in real time.
		In \S\ref{sec:Unroll}, we present an alternative description of the dynamics where we ``unroll" the compact field spaces of the axion and gauge field, exchanging two compact, constrained degrees of freedom for a single noncompact and unconstrained one. We find that there are two ways to do this, which provide two alternative descriptions of the compact dynamics. In the \emph{axion frame}, this single degree of freedom encodes axionic expectation values relatively simply, but the dynamics is non-local---the axion can interact with itself across multiples of its fundamental domain, encoding the original non-trivial topology of its field space. In the \emph{gauge field frame}, this single degree of freedom encodes gauge field expectation values relatively simply, but axionic expectation values are a bit more complicated. However, the dynamics is entirely local. In fact, the Hamiltonian in this frame describes a particle in a periodic potential and a harmonic well, which has been experimentally realized by atomic physicists.  
		
		We then use these different frames to study our class of toy models in two distinct limits. We use the axion frame to study the dynamics of the monodromy-like gapless model in \S\ref{sec:Gaussian}, while in \S\ref{sec:gappedDynamics} we use the gauge field frame to study the instanton-like gapped model. 
		 We will see that the initial ansatz we chose to derive the effective Schr\"{o}dinger equation was nearly the correct one, and we will show how to improve on it. We find that the gauge field frame gives an intuitive picture of both the axion-gauge field dynamics and the origin of the potential correction $V_{0, 0}(\varphi)$.  
		Finally, we conclude in \S\ref{sec:Discussion} with a discussion of our results and promising future directions.

\newpage

	\section{Quantum Mechanical Toy Models} \label{sec:ToyModels} 
		
		Ultimately, we must understand how well a non-perturbatively generated effective potential captures the semi-classical, out-of-equilibrium, dynamics of quantum field theories like (\ref{eq:prototype}). Here, we set our sights a bit lower and instead focus on analogous toy models in quantum mechanics. Similar to how the double-well potential is a useful warm-up for understanding instanton effects in quantum field theory, we are interested in finding the ``double-well'' for theories whose time-evolution is determined entirely by quantum mechanical effects. In this section, we introduce a family of toy models that are structurally similar to (\ref{eq:prototype}) and show that there are indeed stationary solutions to the classical equations of motion that are completely changed by quantum effects.

		There are two essential structures in the theory (\ref{eq:prototype}) that our toy models must mimic:
		\begin{enumerate}
			\item \textbf{ The axion is coupled to the gauge field only through a topological term.}
				
				Continuous shifts of the axion $\phi \to \phi + \epsilon$ transform the Lagrangian (\ref{eq:prototype}) up to a total derivative, and thus do not affect the classical equations of motion. Classically, the axion can sit still anywhere in its field space. Importantly, the action transforms up to a term that depends only on the topology of the gauge field trajectory.

			\item  \textbf{The gauge field's configuration space is topologically non-trivial.}
				
				Many gauge theories are invariant under large gauge transformations $A \to A + {n}\hspace{1pt}{\omega}$, where $n$ is an integer and $\omega$ is a closed, but not exact, form. Ensuring the theory is invariant under these gauge transformations forces the path integral to include a sum over topological sectors \cite{Rajaraman:1982is, Shifman:2012zz}, and the aforementioned topological coupling produces quantum mechanical interference effects that are not present classically.
				
			\end{enumerate}
			 It is the combination of a topologically non-trivial field space and ``total derivative" coupling that generates an effective potential for the axion, and so these are necessary ingredients in any toy model that hopes to imitate (\ref{eq:prototype}). Furthermore, we will also require that:
			\begin{enumerate} \setcounter{enumi}{2}
				\item \textbf{The axion's configuration space is topologically non-trivial.}
				
					In bottom-up models of natural inflation, the axion field space can be either compact or non-compact.\footnote{This is analogous to the choice of gauge group for QED. If the gauge group is compact, then magnetic monopoles are part of the Hilbert space, and their existence generally implies some interesting non-trivial dynamics at long distances, like confinement \cite{Polyakov:1976fu}.} However, in most ultraviolet completions under sound theoretical control, the axion potential is protected from (potentially disastrous) corrections because its field space is a circle---that is, discrete shifts $\phi \to \phi + 2\pi f$ represent a true redundancy of the theory and no correction can violate this gauge symmetry.
			\end{enumerate}
			Though not necessary to generate an effective potential, a compact axionic field space is a generic feature in UV complete models of natural inflation. For instance, if the axion descends from the dimensional reduction of a gauge field, as is common in string compactifications, these discrete shifts are generated by large gauge transformations in the higher-dimensional theory. We are only interested in these types of ``top-down inspired'' theories and so we assume the axion field space is compact. 

	 With these features in mind, we consider a class of toy models described by the Lagrangian
		\begin{equation}
			\mathcal{L} = \frac{f^2}{2} \dot{\varphi}^2 + \frac{1}{2 g^2} \dot{A}^2 - V(A)+ 2\pi \hbar \varphi \dot{A} \label{eq:toyLagrangian}
		\end{equation}
		where both the ``axion'' $\varphi$ and the ``gauge field'' $A$ have compact configuration spaces,
		\begin{equation}
			\varphi \sim \varphi + 1 \qquad \text{and} \qquad A \sim A + 1\,, \label{eq:periods}
		\end{equation}
		and are coupled via a topological interaction. This system describes a particle moving on a torus in the presence of both an electric (the gauge field potential) and magnetic (the topological interaction) field.\footnote{The topological term $k\times 2 \pi \hbar \varphi \dot{A}$ must be quantized ($k \in \mathbbm{Z}$) if the redundancies (\ref{eq:periods}) are to be consistent at the quantum level---we can view this as magnetic flux quantization on the torus. In this paper, we only consider the simplest case, $k=1$.} The gauge field potential $V(A)$ can be thought of as encoding different gauge field dynamics induced by the charged matter fields $\mathcal{L}_\lab{cm}$ in (\ref{eq:prototype}), and will thus control the behavior of the gauge field instantons.
		
		While our results will apply for arbitrary periodic $V(A)$, we will illustrate our points with a single model considered in two limits. The simplest case, described in \S\ref{sec:gaplessPotential}, is to set $V(A) = 0$. We call this the \emph{gapless model}, as the energy gap between the lowest and first excited states of the gauge field closes as the axion winds around its fundamental domain. This model is Gaussian, and the effective potential takes the form of a ``sum over branches,'' familiar from axion monodromy \cite{Dvali:2005an, Silverstein:2008sg,  McAllister:2008hb, Kaloper:2008fb, Kaloper:2011jz, Lawrence:2012ua} and large-$N$ QCD \cite{Witten:1980sp, Witten:1998uka, Halperin:1997bs, Dine:2016sgq}. In \S\ref{sec:gappedPotential}, we study the \emph{gapped model} with potential $V(A) = \Lambda(1 - \cos 2 \pi A)$.  In the large $\Lambda$ limit, we show that the effective potential takes the form of an exponentially-suppressed cosine, calculable by the dilute instanton gas approximation, familiar from natural inflation \cite{Freese:1990rb, Adams:1992bn} and high temperature QCD \cite{Gross:1980br}. While this model assumes a specific form for the potential---one whose properties are easily calculable---we will argue that the conclusions derived from it are universal in the large $\Lambda$ limit.

	\subsection{Classical Considerations} \label{sec:classical}
		We first consider (\ref{eq:toyLagrangian}) at the classical level. The equations of motion are\footnote{We have dropped the topological coupling's dependence on $\hbar$ here, to avoid confusion about what constitutes as ``classical.'' While it is unimportant at the classical level, its $\hbar$-dependence is necessary for the Feynman measure to be invariant under $\varphi \to \varphi +1$.
		}
		\begin{equation}
			f^2 \ddot{\varphi} = 2 \pi  \dot{A} \qquad \text{and} \qquad g^{-2} \ddot{A} + V'(A) + 2 \pi \dot{\varphi} = 0\,. \label{eq:classicalEom}
		\end{equation}
		In the gapless model, $V(A) = 0$, this system executes simple harmonic motion with frequency $\omega = 2 \pi g/f$, while for non-trivial $V(A)$ the motion is more complicated. Importantly, because these equations of motion only depend on time derivatives of $\varphi$, the axion can sit still anywhere along its field space. This will not be the case at the quantum level.
		
		There is a fundamental difference between this system and its higher-dimensional counterpart that we must mention. For simplicity, we will take (\ref{eq:prototype})'s gauge group to be $\lab{U}(1)$, so that it describes classical electrodynamics coupled to an axion with equations of motion
		\begin{gather*}
			\nabla \cdot \mb{B} = 0\,, \qquad\qquad \nabla \times \mb{E} = -\frac{\partial \mb{B}}{\partial t}\,, \qquad \qquad \partial^2 \phi = \frac{g^2}{8 \pi^2 f} \mb{E} \cdot \mb{B}\,,\\
		 \nabla \cdot \mb{E} = - \frac{g^2}{8 \pi^2 f} \left(\nabla \phi\right)\cdot \mb{B}\,,  \qquad \qquad \nabla \times \mb{B} = \frac{\partial \mb{E}}{\partial t} - \frac{g^2}{8 \pi^2 f} \left(\dot{\phi}\, \mb{B} + \left(\nabla \phi\right) \times \mb{E}\right).
	\end{gather*}
		Again, only spacetime derivatives of the axion appear in the equations of motion and so, like our toy model, a spatially constant axion $\phi(t, \mb{x}) = \phi(t)$ may sit still anywhere along its field space. However, since the four-dimensional topological term is quadratic in the gauge field, the axion always appears multiplied either by the electric or magnetic field. If we assume the gauge field sits in its vacuum, i.e. $\mb{E} = \mb{B} = 0$, we see that the axion zero mode may freely evolve in time, $\phi(t) \propto t$. This is in contrast to our toy model, whose topological term is linear in the gauge field. Because we are mainly interested in how classically stationary trajectories are modified by quantum effects, we do not expect\footnote{Another way to view the toy models, where the analogy is more direct, is as the dimensional reduction of an axion coupled to a top-form gauge field, where $F_4 = \ud C_3$, and $C_{3} = \lab{tr}\left(A\, \ud A - 2 A^3/3\right) = \hodge J_\lab{CS}$ \cite{Luscher:1978rn, DiVecchia:1980yfw, Dvali:2005an}. In this case, the coupling of the axion to the topological term is linear in both the axion and the top-form field strength.} that this difference plays an important role. However, we leave a more thorough exploration to future work.

	\subsection{General Quantum Considerations}
		
		While the redundancies (\ref{eq:periods}) do not affect the classical dynamics, they play an important role upon quantization.  The Hamiltonian of our toy model (\ref{eq:toyLagrangian}) is
		\begin{equation}
			\mathcal{H} = \frac{p_\varphi^2}{2 f^2} + \frac{g^2}{2} \left(p_A - 2\pi \hbar \varphi \right)^2 + V(A)\,. \label{eq:Hamiltonian}
		\end{equation}
		The generators of $\varphi$ and $A$ translations are
		\begin{equation}
			\pi_\varphi = p_\varphi - 2\pi \hbar A \qquad \text{and} \qquad \pi_A = p_A\,, \label{eq:generators}
		\end{equation}
		respectively, where $[\varphi, p_\varphi] = [A, p_A] = i \hbar$.\footnote{As is well known from the study of the quantum mechanical angle operator \cite{Barnett:2007qpo}, these canonical commutation relations are modified if $\varphi$ or $A$ are compact. These technical complications are avoidable as long as we work with gauge invariant observables like  (\ref{eq:expValPhi}), and so we only mention them in passing.} Crucially, the topological interaction forces the axionic translation generator $\pi_\varphi$ to also depend on $A$.

		The redundancies (\ref{eq:periods}) impose a gauge constraint on the physical states in the Hilbert space---any translation $\varphi \to \varphi + \ell_\varphi$ or $A \to A + \ell_A$, where $\ell_\varphi$ and $\ell_A$ are integers,  must bring a physical state $|\Psi \rangle$ exactly into itself,
		\begin{equation}
			\exp\left(-i(\pi_\lab{\varphi} \ell_\varphi + \pi_A \ell_A)/\hbar\right) | \Psi \rangle = | \Psi \rangle\,. \label{eq:gaussLaw}
		\end{equation}
		In terms of the wavefunction $\Psi(\varphi, A, t) \equiv \langle \varphi, A| \Psi \rangle$, this constraint imposes  the quasi-periodic boundary conditions
		\begin{equation}
			\exp\left(2\pi i A\right) \Psi(\varphi - 1, A, t) = \Psi(\varphi, A, t) \quad \text{and} \quad \Psi(\varphi, A-1,t) = \Psi(\varphi, A,t)\,. \label{eq:boundaryConditions}
		\end{equation}
		That $\varphi$ and $A$ are linked intrinsically through these boundary conditions will play an enormous role in the following analysis. It will allow us to effectively reduce the theory to that of a single non-compact degree of freedom, and determine its quantum dynamics exactly.

		\begin{figure}
			\centering
			\includegraphics[scale=0.95, trim = 0 0 0 0]{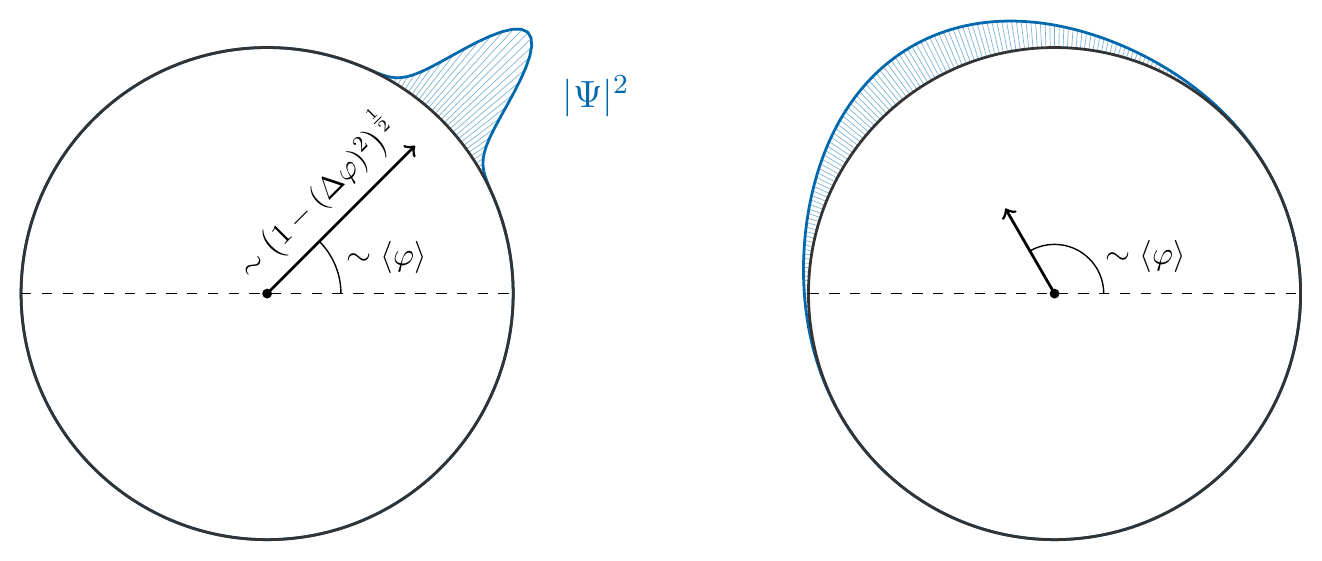}
			\caption{
			We must define the axion's position through a gauge-invariant observable, as gauge-variant expectation values like $\langle \varphi^n\rangle$ are difficult to interpret. A natural choice, illustrated above, is $\langle e^{2 \pi i \varphi}\rangle_\Psi$, whose phase and magnitude roughly measure the position and spread of the axion, respectively. \label{fig:expectationValues}}
		\end{figure}
		
		Observables of the theory must be gauge invariant expectation values. For example, in order to measure the expectation value of $\varphi$, i.e. its position, we must instead compute
		\begin{equation}
			\langle e^{2 \pi i m \varphi} \rangle_\Psi = \int_{\sminus \frac{1}{2}}^{\frac{1}{2}}\!\ud \varphi \int_{\sminus\frac{1}{2}}^{\frac{1}{2}} \!\ud A\,  e^{2 \pi i m \varphi} \left|\Psi(\varphi, A, t)\right|^2, \qquad m \in \mathbbm{Z}. \label{eq:expValPhi}
		\end{equation}
		The expectation value $\langle e^{2 \pi i \varphi} \rangle_\Psi$ (see Fig. \ref{fig:expectationValues}) contains information about both the position of the axion and how well-localized it is---roughly, 
			\begin{equation}
				\langle e^{2 \pi i \varphi} \rangle_\Psi \sim \big[1 - (\Delta \varphi)^2\big]^{1/2} \exp\left(2 \pi i \langle \varphi \rangle\right).
			\end{equation}
			
			Finally, we should note that the boundary conditions (\ref{eq:boundaryConditions}) can be modified by an arbitrary phase that corresponds to an inequivalent quantization of the theory---in practice, this amounts to adding a $\theta$-angle to either the gauge field or the axion.\footnote{While an additional $\theta$-angle for the gauge field can be absorbed by a shift of the axion $\varphi \to \varphi -\theta/2\pi$, a $\theta$-angle for the axion would change the very low-energy dynamics of the theory. We leave an exploration of this to future work.}

	\subsection{Effective Potential and the Instanton Expansion} \label{sec:effPotInst}
		
		Our toy models must pass one key test if they are to mimic the higher dimensional model (\ref{eq:prototype}): integrating out the gauge field $A$ must generate an effective potential for $\varphi$.
		As we discussed in \S\ref{sec:classical}, classically the axion can sit still anywhere along its field space, so that the effective potential is naively zero due to the shift symmetry generated by $\pi_\varphi$ (\ref{eq:generators}). However, this continuous shift symmetry is ``broken'' to a discrete shift symmetry once the periodic boundary conditions (\ref{eq:boundaryConditions}) are imposed---only the discrete shifts generated by $\exp\left(-i \pi_\varphi \ell /\hbar\right)$, with integer $\ell$, are symmetries. At the level of the path integral, the gauge field's compact field space forces one to sum over topological sectors, and interference effects among these different sectors generate an effective potential upon integrating out $A$.
		
		We can compute this potential in two ways, both of which assume that the axion is a fixed background field and that the gauge field rests in its vacuum state.  The axion is then expected to evolve adiabatically, slowly enough that the gauge field remains in its vacuum. We will see that this assumption imposes ``in'' and ``out'' boundary conditions on the gauge field that do not hold in out-of-equilibrium dynamics. In \S\ref{sec:effSchrodinger}, we relax this assumption to quantify the corrections that appear.

		The first method is to find the ground state energy from the gauge field Schr\"{o}dinger equation, as a function of the now classical parameter $\varphi$.
	This is equivalent to studying the spectrum of the Hamiltonian
		\begin{equation}
			\mathcal{H}_\lab{c} = \frac{g^2 p_A^2}{2} + V(A)
		\end{equation}
		subject to the axion-dependent boundary condition $\psi(A+1) = e^{\sminus 2 \pi i \varphi} \psi(A)$. Ignoring this boundary condition for the moment, this is simply the Hamiltonian for a particle of mass $g^{-2}$ propagating in an infinite periodic potential, i.e. a particle in a one-dimensional crystal lattice. It is thus natural to utilize the technology developed in condensed matter physics \cite{ashcroft1976solid,chaikin1995principles} to study these systems. By Bloch's theorem, the energy eigenstates of $\mathcal{H}_\lab{c}$ are the quasi-periodic Bloch waves,
		\begin{equation}
			\mathcal{H}_\lab{c} \psi_{n, \kappa}(A) = E_{n,\kappa}\psi_{n,\kappa}(A) \qquad \text{with} \qquad \psi_{n, \kappa} (A+1) = e^{2 \pi i \kappa} \psi_{n, \kappa}(A)\,. \label{eq:bloch}
		\end{equation}
		The energies $E_{n}(\kappa) \equiv E_{n, \kappa}$ are arranged in bands (see Figure \ref{fig:mathieuEnergies}) and are labeled by a non-negative integer band number $n$ and a ``crystal momentum'' $\kappa \in [\minus 1/2, 1/2)$. Including the boundary condition then fixes~$\kappa = \minus \varphi$, so that the effective potential is simply the lowest energy band,
		\begin{equation}
			V_\lab{eff}(\varphi) = E_{0}(\minus \varphi)\,. \label{eq:effPotGen}
		\end{equation}
		This definition of the effective potential highlights the fact that both the gauge field and axion are irrevocably linked---the axion fixes the gauge field's boundary conditions.  Throughout some putative axion dynamics, evolution in the effective potential assumes that the gauge field adibatically evolves, always sitting in its instantaneous vacuum state. 
		
		An alternative (albeit more familiar to particle physicists) definition of the effective potential is via the Euclidean path integral
		\begin{equation}
			V_\lab{eff}(\varphi) \equiv  \lim_{\beta \to \infty} -\frac{\hbar}{\beta} \log \mathcal{Z}(\beta, \varphi)
		\end{equation}
		where we define
 		\begin{equation}
 			\mathcal{Z}(\beta, \varphi) =  \int\!\mathcal{D} A\, \exp\left[-\frac{1}{\hbar} \int_{\sminus \beta/2}^{\beta/2}\!\ud \tau\, \left(\frac{1}{2 g^2} \dot{A}^2 + V(A)\right) + 2\pi i \varphi \int_{\sminus \beta/2}^{\beta/2}\!\ud \tau \dot{A}\right]
 		\end{equation}
 		and slightly abuse notation to write ${\dot{A} = \ud A/\ud \tau}$, where $\tau$ is Euclidean time.
 		
 		We will now review the computation of the effective potential using these two methods for both the gapless and gapped models.

	\subsubsection{Gapless Model} \label{sec:gaplessPotential}
			We first focus on the Gaussian gapless model. Setting $V(A) = 0$, the energy eigenstates of $\mathcal{H}_\lab{c}$ are the simple Fourier modes
			\begin{equation}
				\psi_{n, \sminus \varphi}(A) = \exp\left(2 \pi i (n - \varphi + \lfloor \varphi \rceil) A\right)
			\end{equation}
			with energies
			\begin{equation}
				E_{n}(\minus \varphi) = \frac{(2 \pi \hbar g)^2}{2} \left(n - \varphi+\lfloor \varphi \rceil\right)^2.
			\end{equation}
			The effective potential is then the lowest energy band,
			\begin{equation}
				V_\lab{eff}(\varphi) = E_{0}(\minus \varphi) = \frac{(2 \pi \hbar g)^2}{2} \left(\varphi - \lfloor \varphi \rceil\right)^2. \label{eq:effPotentialMonodromy}
			\end{equation}
			We must explain the dependence on $\varphi$. By convention, dragging the axion across the boundary of its fundamental domain does not change the energy level $n$. The Bloch waves can then only depend on the combination $\varphi - \lfloor \varphi \rceil$, where $\lfloor \varphi \rceil$ is the nearest integer function,\footnote{For example, $\lfloor 1.25 \rceil=1$, $\lfloor 1.5 \rceil=1$, and $\lfloor 1.7 \rceil=2$.} and are periodic in $\varphi$.

			\begin{figure}
				\begin{center}
					\hspace{-40pt}\includegraphics[scale=1, trim = 0 0 0 0]{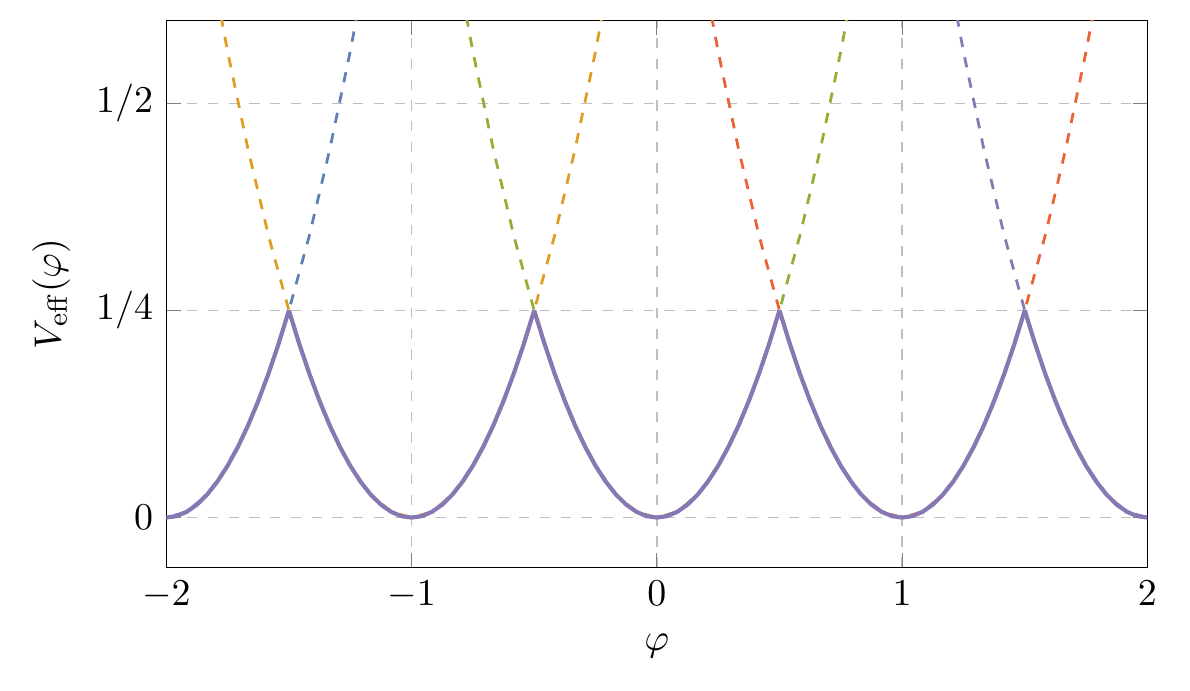}
					\caption{The gapless, ``monodromy-like,'' effective potential (solid, in units of $(2 \pi \hbar g)^2/2$), is the minimum of multiple quadratic branches (dashed). The effective potential predicts that the axion will execute simple harmonic motion for small energies, while for high energies it will wind around the axion's compact field space indefinitely.}\label{fig:monodromyPotential}
				\end{center}
			\end{figure}

			We can also compute the effective potential by considering the low temperature $\beta \to \infty$ limit of the Euclidean path integral,
			\begin{equation}
				\mathcal{Z}(\beta) = \int\limits_{\mathrlap{A(0) \sim A(\beta)}}\!  \mathcal{D} A(\tau)\, \exp\left(-\frac{1}{2 g^2 \hbar} \int_{\sminus \beta/2}^{\beta/2} \!\!\ud \tau\, \dot{A}^2 + 2\pi i \int_{\sminus \beta/2}^{\beta/2}\!\!\ud \tau\, \varphi\, \dot{A}\right).
			\end{equation}
			 This integral is over all closed paths on the circle $A \sim A+1$, which we may instead represent as a sum over topological sectors
			\begin{equation}
				\mathcal{Z}(\beta) = \sum_{\ell \in \mathbbm{Z}}  \int_{0}^{1} \!\ud A_0\, \int\limits_{\mathrlap{A_0}}^{\mathrlap{A_0 + \ell}}  \mathcal{D} A(\tau)\, \exp\left(-\frac{1}{2 g^2 \hbar} \int_{\sminus \beta/2}^{\beta/2}\!\!\!\ud \tau\, \dot{A}^2 + 2\pi i \int_{\sminus \beta/2}^{\beta/2}\!\!\!\ud \tau\, \varphi \,\dot{A}\right)
			\end{equation}
			of paths on the real line. For each term in the sum, we shift $A(\tau) \to \bar{A}_\ell(\tau) + a(\tau)$ where
			\begin{equation}
				\bar{A}_\ell(\tau) = A_0 +  \frac{\ell}{\beta} \left(\tau + \frac{\beta}{2}\right) \qquad\, \text{with} \qquad  a(0) = a(\beta) = 0\, \label{eq:tunnelingConfigs}
			\end{equation}
			is the classical path connecting $A(0) = A_0$ and $A(\beta) = A_0 + \ell$.
			
			The path integral can then be written as
			\begin{equation}
				\mathcal{Z}(\beta) =  \mathcal{K}[\varphi(\tau)] \sum_{\ell \in \mathbbm{Z}} \exp\left(-\frac{\ell^2}{2 g^2 \hbar \beta} + 2\pi i \ell \bar{\varphi} \right).  \label{eq:monodromyPartitionFunction}
			\end{equation}
			Here, we have defined the axion's average position, $\bar{\varphi} = \beta^{-1} \int_{\sminus \beta/2}^{\beta/2} \! \ud\tau \, \varphi(\tau)$,
			and the fluctuation determinant
			\begin{equation}
				\mathcal{K}[\varphi(\tau)] \equiv \int\limits_{\mathrlap{a(\sminus \beta/2) = 0}}^{\mathrlap{a(\beta/2) = 0}}\!  \mathcal{D} a(\tau)\,\exp\left(-\frac{1}{2 g^2 \hbar} \int_{\sminus \beta/2}^{\beta/2}\!\!\ud \tau\, \dot{a}^2 - 2 \pi i \int_{\sminus \beta/2}^{\beta/2}\!\!\ud \tau\, \dot{\varphi} \,a\right).
			\end{equation}
			
			Transitions between the different topological sectors are mediated by the classical trajectories (\ref{eq:tunnelingConfigs}), with Euclidean action
			\begin{equation}
				S_\lab{E, \ell} = \frac{\ell^2}{2 g^2 \beta}\,.
			\end{equation}
			Depending on one's tastes, we might call these ``instantons'' though they are not at all localized at a single instant in time. In fact, in the $\beta \to \infty$ limit, these trajectories become arbitrarily delocalized, their actions all approach zero, and the dilute instanton gas approximation is not reliable or appropriate. Fortunately, this model is Gaussian so we need not rely on this approximation.
			
			For constant $\varphi(\tau) = \bar{\varphi}$, the fluctuation determinant (\ref{eq:monodromyPartitionFunction}) is a constant coefficient. Ignoring this constant factor, we may use Poisson resummation to reexpress  (\ref{eq:monodromyPartitionFunction}) as
			\begin{equation}
				\log \mathcal{Z}(\beta) 
				\sim \log \left[\sum_{\ell \in \mathbbm{Z}} \exp\left(\minus \frac{(2 \pi \hbar g)^2 \beta}{2 \hbar} \left( \varphi - \ell \right)^2 \right)\right],
			\end{equation}
			which is dominated by the $\ell = \lfloor \varphi \rceil$ term in the $\beta \to \infty$ limit. We thus reproduce the effective potential derived via canonical methods, 
			\begin{equation}
				V_\lab{eff}(\varphi) = \frac{(2 \hbar g)^2}{2} \sum_{n = 1}^{\infty} \frac{(-1)^{n+1}}{n^2} \left(1 - \cos 2 \pi n \varphi\right) = \frac{(2 \pi \hbar g)^2}{2} \left(\varphi - \lfloor \varphi \rceil \right)^2.
			\end{equation}
			
			As shown in Figure \ref{fig:monodromyPotential}, the effective potential can be thought of as a ``sum over branches,'' or is ``monodromy-like'' in the language of \cite{Dine:2016sgq}. These types of potentials have been studied in the context of axion monodromy \cite{Lawrence:2012ua, McAllister:2008hb} and large-$N$ QCD \cite{Witten:1980sp, Witten:1998uka, Halperin:1997bs}, and we take this gapless model as a low-dimensional avatar of these higher-dimensional models.

	\subsubsection{Gapped Model} \label{sec:gappedPotential}
			Many models of inflation rely on an effective potential of the form predicted by the dilute instanton gas approximation,
			\begin{equation}
				V_\lab{eff}(\varphi) \sim \minus\Lambda\, e^{-\mathcal{S}_1} \cos 2 \pi \varphi\,,
			\end{equation}
			where $\mathcal{S}_1$ is the dimensionless one-instanton action. Since we are primarily interested in making contact with these applications, the gapless model suffers a major drawback---there is no ``energy'' cost for $A$ to wind around its field space arbitrarily slowly and thus there is no sense in which the instantons are dilute.  In order to avoid this behavior, we may localize the instantons by including a potential for $A$. In what follows, we will consider the simple cosine potential $V(A) = \Lambda(1 - \cos 2 \pi A)$, which has been studied in a number of contexts \cite{Slater:1952asp, Asorey:1983hd}.

 		 	\begin{figure}
				\begin{center}
					\hspace{-5pt}\includegraphics[scale=1, trim = 0 0 0 0]{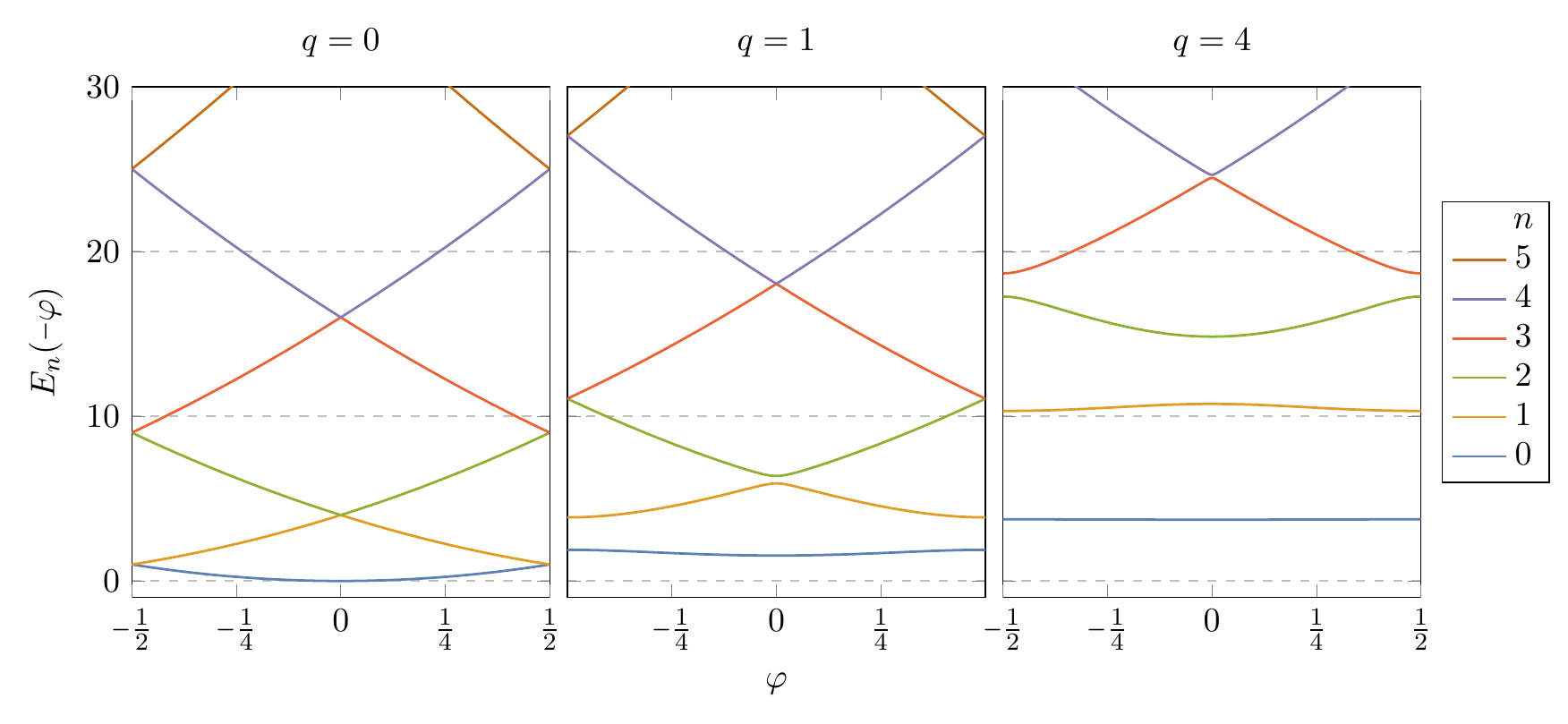}
					\caption{Energies of the various bands as a function of the gauge field ``mass gap'' $\Lambda$, quantified by $q\equiv \Lambda/(\pi^2 g^2 \hbar^2)$. As $q$ increases, the gaps between the various bands appear and grow. For large $q$, the lowest energy band becomes ``instanton-like,'' and is roughly a single cosine with exponentially suppressed amplitude. 
					\label{fig:mathieuEnergies}}
				\end{center}
		\end{figure}

		As before, we start with the Hamiltonian approach. The energy eigenvalues are determined by the Mathieu equation\footnote{We explain our conventions for the Mathieu functions in \S\ref{app:mathieu}.}
		\begin{equation}
			\left(\frac{\hbar^2 g^2}{2} \partial_A^2  + E_{n}(\kappa) - \Lambda (1 - \cos 2 \pi A) \right)\psi_{n, \kappa} = 0\,,
		\end{equation}
		whose quasi-periodic eigenfunctions are denoted as
		\begin{equation}
			\psi_{n, \kappa}(A) \propto \lab{me}_{2 n + 2\kappa}(\pi A, \minus q)\,,
		\end{equation}
		where $q = \Lambda/(\pi^2 g^2 \hbar^2)$.  The energies are determined by the Mathieu eigenvalues $\lambda_{\nu}$,
		\begin{equation}
			E_{n}(\kappa) = \frac{\pi^2 g^2 \hbar^2}{2} \Big(2 q + 	\lambda_{2n+2\kappa} ( \minus q)\Big)\,, \label{eq:exactEnergyMathieu}
		\end{equation}
		so that for $|\varphi| \leq 1/2$ the effective potential is
		\begin{equation}
				V_\lab{eff}(\varphi) 
				= E_{0}(\minus \varphi)\,.
		\end{equation}
		In Figure \ref{fig:mathieuEnergies}, we plot the lowest lying energy bands for a few values of $q$. When the potential $V(A)$ vanishes, we recover the gapless model and the different energy bands intersect at $\varphi = 0$ and $\pm 1/2$. As $q$ increases, an energy gap $E_{1}(1/2) - E_{0}(1/2) \sim 2 \pi^2 g^2 \hbar^2 \sqrt{q}$ appears \cite{Wilkinson:2018asm} and the lowest lying band is well-described by an exponentially suppressed cosine,
		\begin{equation}
			V_\lab{eff}(\varphi) \sim \frac{\pi^2 g^2 \hbar^2}{2}\left(2 \sqrt{q} -\frac{1}{4}  - \frac{32}{\sqrt{2\pi}} q^{3/4} e^{-4 \sqrt{q}} \cos 2\pi \varphi + \dots\right) \quad \text{as} \quad q \to \infty\,.
		\end{equation}
		This behavior of the energy bands is the origin of our names for the gapless and gapped models.

			We may also compute this using the Euclidean path integral
			\begin{equation}
				\mathcal{Z}(\beta, \varphi) = \int\!\mathcal{D} A\, \exp\left(-\frac{1}{\hbar} \int_{\sminus \beta/2}^{\beta/2} \!\!\ud \tau\, \left[\frac{1}{2 g^2} \dot{A}^2 + \Lambda\big(1 - \cos(2 \pi A)\big)\right] + 2\pi i  \varphi\int_{\sminus \beta/2}^{\beta/2}\!\!\ud \tau\,  \dot{A}\right),
			\end{equation}
			where we keep $\varphi$ a classical parameter and integrate over paths that start at $A(\minus \beta/2) = 0$ and end at those that are equivalent to this point, $A(\beta/2) \sim 0$. As before, we may rewrite this as a sum over topological sectors,
			\begin{equation}
				\mathcal{Z}(\beta, \varphi) = \sum_{n \in \mathbbm{Z}} e^{2\pi i n \varphi } \int_{0}^{n}\!  \mathcal{D} A\, \exp\left(-\frac{1}{\hbar} \int_{\sminus \beta/2}^{\beta/2} \!\!\ud \tau \, \left[\frac{1}{2 g^2} \dot{A}^2 + \Lambda \left(1- \cos(2 \pi A)\right)\right]\right).
			\end{equation}
			In the dilute instanton gas approximation \cite{Asorey:1983hd}, the path integral is dominated, as $\beta \to \infty$, by the \mbox{$(\pm 1)$-instantons}
			\begin{equation}
				\bar{A}_{\pm 1} (t) = \frac{2}{\pi}\arctan\left(\exp\left[\pm 2\pi g \sqrt{\Lambda} (\tau-\tau_0)\right]\right). \label{eq:mathieuInstanton}
			\end{equation}
			Unlike in the previous model, these instanton configurations are localized in time, with width
			\begin{equation}
				\Delta t \sim \frac{1}{2 \pi g \sqrt{\Lambda}}\,,
			\end{equation}
			with $\beta$-independent action 
			\begin{equation}
					S_{\pm 1} = \frac{4 \sqrt{\Lambda}}{\pi g} = 4 \hbar \sqrt{q}\,.
			\end{equation}
			The Euclidean path integral is then well approximated by
			\begin{equation}
				\mathcal{Z}(\beta, \varphi) \underset{\beta \to \infty}{\sim} \mathcal{K}_0\sum_{\ell, \bar{\ell} \in \mathbbm{Z}}   \frac{\left(\beta \mathcal{K}_1\right)^{\ell+\bar{\ell}}}{\ell! \bar{\ell}!}e^{-4 \sqrt{q}(\ell+\bar{\ell}) +i 2\pi (\ell-\bar{\ell})\varphi} = \mathcal{K}_0 \exp\left(2 \beta \mathcal{K}_1 e^{-4 \sqrt{q}} \cos 2\pi \varphi\right)
			\end{equation}
			where $\mathcal{K}_0$ and $\mathcal{K}_1$ are the contributions coming from integrating over fluctuations about the minimum at $A = 0$ and the single (anti)-instanton configuration (\ref{eq:mathieuInstanton}), respectively. 
			Both of these contributions can be evaluated via standard techniques, yielding \cite{Asorey:1983hd}
			\begin{equation}
				V_\lab{eff}(\varphi) = \lim_{\beta \to \infty} -\frac{\hbar}{\beta} \log \mathcal{Z}  \sim \frac{\pi^2 g^2 \hbar^2}{2} \left(2 \sqrt{q}  - \frac{32}{\sqrt{2\pi}} q^{3/4} e^{-4 \sqrt{q}} \cos 2\pi \varphi + \dots\right)\,.
			\end{equation}
			
			In the language of \cite{Dine:2016sgq}, these potentials are ``instanton-like'' and we consider these gapped models as low-dimensional avatars of natural inflation UV completions.

	\subsection{Recap}
	
		Our ultimate goal is to quantify the time-dependent corrections to non-perturbatively generated effective potentials in quantum field theories like (\ref{eq:prototype}). Said differently, how do we consistently integrate out non-perturbative effects in time-dependent settings? To make progress, we have introduced a class of toy quantum mechanical models that, we argue, capture the relevant features of our prototype and have shown that they pass a key test---gauge field instantons generate an effective potential for the axion. These toy models thus serve as a test bed for the dynamics of an axion zero mode in quantum field theory. 
		
		In fact, as we explain in Appendix \ref{app:qft}, these toy models can be seen as ``minisuperspace'' truncations of simple quantum field theories in higher dimensions. While there may be additional effects that appear in quantum field theory that are not captured by these low-dimensional models, our goal is to understand the dynamical consequences of the three minimal ingredients outlined in the beginning of this section. Still, since we are interested in the dynamics of a quantum field's \emph{zero mode}---which behaves as a single degree of freedom---we expect the lessons learned here will be valuable in the transition to higher dimensions. 
		Whatever the relation, we leave this question to be settled in future work, and move on to investigate the low-energy dynamics of our toy models.

	\section{Effective Schr\"{o}dinger Equation} \label{sec:effSchrodinger}

			In this section, we introduce an effective Schr\"{o}dinger equation for our toy model that makes its low-energy dynamics manifest.  This description will explicitly demonstrate that the axion sees more than the effective potential, and we will be able to quantify these out-of-equilibrium corrections. Later, in \S\ref{sec:gappedDynamics}, we derive this effective Schr\"{o}dinger equation from a different perspective, putting it on a firmer conceptual footing. First, however, we must describe both what we mean by ``low-energy dynamics'' and motivate why an effective Schr\"{o}dinger equation is a useful way of packaging this information.
			
			As described in the previous section, the effective potential measures the \emph{vacuum energy} of the gauge field as a function of a fixed classical parameter. Once we promote the axion from fixed parameter to dynamical degree of freedom, the hope is that it is allowed to evolve in time while the gauge field remains fixed in its instantaneous vacuum state, so that the dynamic axion ``sees'' the effective potential. Of course, this is too much to ask for. It is difficult, if not impossible, for the axion to evolve without disturbing the gauge field---they are coupled! We are not necessarily interested in understanding the dynamics of the lowest energy states in the full axion-gauge field system but, instead, the set of states where the gauge field is, in some sense, ``close'' to its~vacuum~state. 
		
		Typically, quantum-corrected equations of motion for in-in expectation values like $\langle \varphi(t) \rangle$ are derived via the tadpole method \cite{Jordan:1986ug, Calzetta:1986cq, Traschen:1990sw, Boyanovsky:1994me, Baacke:1996se, Cametti:1999ii, Mooij:2011fi}. Instead, we choose to phrase the dynamics in terms of an effective Schr\"{o}dinger equation for two main reasons. First, because the axion has a compact field space, the expectation value $\langle \varphi(t) \rangle$ is not gauge-invariant. Gauge-invariant expectation values like $\langle \exp(2 \pi i \varphi(t)) \rangle$ will typically satisfy complicated equations of motion, even if the dynamics are relatively simple, and so it is not clear that this is a fruitful direction. Fortunately, the Schr\"{o}dinger equation has no problem with using gauge-dependent variables. Instead, it is our job to restrict ourselves to proper observables in interpreting the resulting wavefunction.
		
		Second, effective quantum dynamics can be extremely state-dependent and it can be difficult to interpret the actual dynamics if the wavefunction is allowed to spread. This is particularly dangerous when trying to make the connection to quantum field theory, as the zero modes of quantum fields behave classically.\footnote{We thank Nima Arkani-Hamed for raising this point.} Working with an effective Schr\"{o}dinger equation is thus helpful at the level of interpretation, as we can understand the dynamics without constraining ourselves to a particular axionic initial state, all the while remaining close to a particular gauge field state.
		
		We note that, at first, this effective Schr\"{o}dinger equation will be nothing more than an alternative representation of the full Schr\"{o}dinger equation, as it amounts to decomposing the general state $\Psi(\varphi, A, t)$ using a particular basis of functions. Ideally, our choice of basis---alternatively, our choice of initial state---makes it obvious which degrees of freedom to ignore in order to find a simplified description. The goal, after all, is to derive an effective description of the axion dynamics that allows us to ignore the gauge field's time evolution entirely. To understand when this description breaks down, we can then compare the dynamics of the effective Schr\"{o}dinger equation before and after truncation. The axion's dynamics will depend sensitively on our choice of basis, and our goal is to choose the correct basis that describes the low-energy dynamics of the full theory. In this section, we decompose the wavefunction in terms of the naive choice---the gauge field's eigenfunctions---and show that the effective description breaks down rather quickly. Still, it will be a useful exercise and in \S\ref{sec:gappedDynamics} we show how to improve upon this choice to derive a more accurate effective description.

	 	We begin by writing the full wavefunction as a sum over products of axionic wavefunctions $\Phi_{n}$ and gauge field eigenfunctions $\psi_{n, \sminus \varphi}$ (\ref{eq:bloch}),
		\begin{equation}
				\Psi(\varphi, A, t) = \sum_{n = 0}^{\infty} \Phi_{n}(\varphi, t) e^{2 \pi i \varphi A} \psi_{n, \sminus\varphi}(A)\,. \label{eq:ansatz}
		\end{equation}
		The $\psi_{n, \sminus \varphi}$ provide a complete basis set for (quasi)-periodic functions in $A$, so this decomposition is unique,
		\begin{equation}
			\Phi_{n}(\varphi, t) = \int_{\sminus 1/2}^{1/2}\!\ud A\, e^{\sminus 2 \pi i \varphi A} \bar{\psi}_{n, \sminus \varphi}(A) \Psi(\varphi, A, t)\,, \label{eq:orthonormality}
		\end{equation}
		and axionic expectation values (\ref{eq:expValPhi}) can be computed using a modified Born rule,
		 \begin{equation}
		 	\langle e^{2 \pi i m \varphi} \rangle_{\Psi} = \sum_{n = 0}^{\infty} \int_{\sminus 1/2}^{1/2}\!\ud \varphi\, e^{2 \pi i m \varphi} |\Phi_{n}(\varphi, t)|^2 \, . \label{eq:modBornRule}
		 \end{equation}
		 The boundary conditions (\ref{eq:boundaryConditions}) imply that each of the axionic wavefunctions are strictly periodic, $\Phi_{n}(\varphi-1, t) = \Phi_{n}(\varphi, t)$, and so it is very tempting to identify $\Phi_{0}(\varphi, t)$ as the wavefunction of the axion when the gauge field has been constrained to its vacuum. We will see in \S\ref{sec:gappedDynamics} that this interpretation is correct.

		 Inserting this expansion into the full Schr\"{o}dinger equation and using (\ref{eq:orthonormality}), we may derive a set of coupled, effective Schr\"{o}dinger equations for the $\Phi_n$, 
			\begin{equation}
				\left(-i \hbar \, \partial_t + \frac{p_\varphi^2}{2 f^2}  + E_{n}(\minus \varphi)\right) \Phi_{n} + \sum_{n' = 0}^{\infty}  \left[ F_{n, n'}(\varphi) p_{\varphi} \Phi_{n'} + V_{n, n'}(\varphi) \Phi_{n'}\right] = 0\,,
			\end{equation}
			where the terms
			\begin{align}
				F_{n, n'} &\equiv \frac{1}{f^2} \int_{\sminus 1/2}^{1/2}\!\!\ud A\, \left(e^{\sminus 2 \pi i \varphi A} \bar{\psi}_{n, \sminus \varphi}\right) p_{\varphi}\left(e^{2 \pi i \varphi A} \psi_{n', \sminus \varphi}\right) = \langle n, \minus\varphi| e^{\sminus 2 \pi i \varphi A} p_\varphi e^{2 \pi i \varphi A} | n', \minus\varphi \rangle\label{eq:fnn}
			\end{align}
			and
			\begin{align}
				V_{n, n'} &\equiv \frac{1}{2 f^2} \int_{\sminus 1/2}^{1/2}\!\!\ud A\, \left(e^{\sminus 2 \pi i \varphi A} \bar{\psi}_{n, \sminus \varphi}\right) p^2_{\varphi}\left(e^{2 \pi i \varphi A} \psi_{n', \sminus \varphi}\right) = \langle n, \minus\varphi | e^{\sminus 2 \pi i \varphi A} p_\varphi^2 e^{2 \pi i \varphi A} | n', \minus\varphi\rangle \label{eq:vnn}
			\end{align}
			arise from a simple application of the chain rule, i.e. $p_\varphi = \minus i \hbar \partial_\varphi$. Here, we have written $\langle A | n, \minus\varphi\rangle \equiv \psi_{n, \sminus \varphi}(A)$. It will be useful to rewrite this as
			\begin{align}
				i \hbar\, \partial_t \Phi_{n} &= \left(\frac{1}{2 f^2} \left(p_\varphi + f^2 F_{n,n}\right)^2 + E_{n} + V_{n, n} - \frac{1}{2} \left(p_\varphi F_{n, n}\right) - \frac{1}{2} f^2 F_{n, n}^2\right) \Phi_{n} \nonumber \\
				&+ \sum_{n' \neq n} \left[F_{n, n'}(p_\varphi + f^2 F_{n',n'}) + V_{n,n'} - f^2 F_{n, n'} F_{n', n'}\right]\Phi_{n'}\,,
			\end{align}
			where both the combination 
			\begin{equation}
				V_{n, n} - \frac{1}{2}\left(p_\varphi F_{n, n}\right) = \frac{1}{2 f^2} \int_{\sminus 1/2}^{1/2}\!\ud A\, \big|p_\varphi\left(e^{2 \pi i \varphi A} \psi_{n, \sminus \varphi}\right)\!\big|^2
			\end{equation}
			and the $F_{n, n}$ are manifestly real. 
			
			Now, there is an ambiguity in (\ref{eq:ansatz}) that we must address. The Bloch wave functions are determined by the Schr\"{o}dinger equation (\ref{eq:bloch}) up to an overall, possibly $\varphi$-dependent, phase. Changing this phase is degenerate with the gauge transformations $\Phi_{n} \to e^{i \chi_n(\varphi)} \Phi_{n}$ and does not affect physical observables. We may thus always choose a $\chi_{n}(\varphi)$ to completely remove the potentials $F_{n, n}$  so that, concentrating on the $n=0$ sector, we are left with the effective Schr\"{o}dinger equation
			\begin{align}
				i \hbar\, \partial_t \Phi_{{ 0}} &= \left(\frac{p_\varphi^2}{2 f^2}   + V_\lab{eff}(\varphi) + V_{{0,} {0}}(\varphi)\right) \Phi_{0}  + \sum_{n' \neq { 0}} \left[F_{{0}, n'}(\varphi) \, p_\varphi  + V_{0,n'}(\varphi) \right]\Phi_{n'}\,. \label{eq:effEq}
			\end{align}

			We expect that this $n=0$ sector describes the dynamics of the system when the gauge field is close to its vacuum state and we will confirm this intuition in \S\ref{sec:gappedDynamics}. We find that the axionic wavefunction $\Phi_0$ ``sees'' the effective potential $V_\lab{eff}(\varphi) = E_{0}(\minus \varphi)$ and two additional effects: a correction to the effective potential and friction-like couplings to the other axionic wavefunctions~$\Phi_n$.  
			
			The correction to the effective potential is a bit surprising and is due, in some sense, to the quantum mechanical fluctuations of the axion, as it cannot be seen if we treat the axion as a fixed classical parameter. However, it does not depend on how delocalized the axion is. Typically, this sort of correction would disappear in the semi-classical limit. However, because it is competing with a potential of quantum mechanical origin, it can be sizable and important.
			The friction terms $F_{n, n'}$ and $V_{n, n'}$ are expected---the axion is coupled to the gauge field, so it should be able to dump energy into higher gauge field excitations. The $V_{n,n'}$ represent a failure of our effective description, as they encode the velocity-independent leakage of probability into the excited states of the gauge field. We will see later that these terms can be suppressed by a more refined choice of basis. The $F_{n, n'}$ cannot be suppressed by a choice of basis, and represent a speeding axion's ability to drive the gauge field into an excited state.
			
			 If we can drop these friction-like couplings, we arrive at an effective description in which the axion evolves in the potential $V_\lab{eff}(\varphi) + V_{0, 0}(\varphi)$, and axionic expectation values are computed with $\Phi_0$ using the usual Born rule. Now, we compute these potentials in both the gapless and gapped models.

			\subsection{Gapless Model} \label{sec:gaplessPotentials}
				
			The Bloch waves for the gapless model are the simple plane waves,
				\begin{equation}
					\psi_{n, \sminus \varphi}(A) = \exp\big(2 \pi i (n - \varphi + \lfloor \varphi \rceil) A\big)\,, \label{eq:gaplessBloch2}
				\end{equation}
				and we may evaluate the potentials $F_{n, n'}$ and $V_{n, n'}$ explicitly,
				\begin{align}
					F_{n, n'} &=
					 \begin{dcases} \frac{i \hbar}{f^2} \frac{(-1)^{n+n'}}{n - n'} \partial_\varphi \lfloor \varphi \rceil &  n \neq n'  \\
					0 & n = n'
					\end{dcases} \label{eq:gaplessFnn}\\
					V_{n, n'} 
					&= \begin{dcases} \frac{(-1)^{n-n'}\hbar^2}{f^2} \left(\frac{\partial_\varphi^2 \lfloor \varphi \rceil }{2(n - n')} + \left(\frac{ \partial_\varphi \lfloor \varphi \rceil}{n-n'}\right)^2\right) & n \neq n'				 \\
					 -\frac{1}{12}\left(\frac{2 \pi \hbar}{f}\right)^2 \left(\partial_\varphi \lfloor \varphi \rceil\right)^2 & n = n'
					 \end{dcases} \label{eq:gaplessVnn}
				\end{align}
				where
				\begin{equation}
					\partial_\varphi \lfloor \varphi \rceil = \sum_{k \in \mathbbm{Z}} \delta\!\left(\varphi - k + \tfrac{1}{2}\right).
				\end{equation}
				
				These corrections vanish everywhere except at the edges of the axion's fundamental domain, $\varphi = k + 1/2$ for integer $k$. As long as the axion does not cross these points, the effective Schr\"{o}dinger equation predicts that it will undergo simple harmonic motion. However, what happens if the axion has enough energy to reach, say, $\varphi = 1/2$? Apparently something drastic, though the dynamics is tricky to analyze in this picture. In \S\ref{sec:Gaussian}, we will find a more elegant way of studying this system, and show that this singular behavior is simply encoding the fact that the axion sees a full harmonic potential and not the ``cuspy'' effective potential (\ref{eq:effPotentialMonodromy}). It thus allows for simple harmonic motion of arbitrary amplitude.

			\subsection{Gapped Model} \label{sec:gappedPotentials}

				\begin{figure}
					\hspace{-30pt}
					\includegraphics[scale=1]{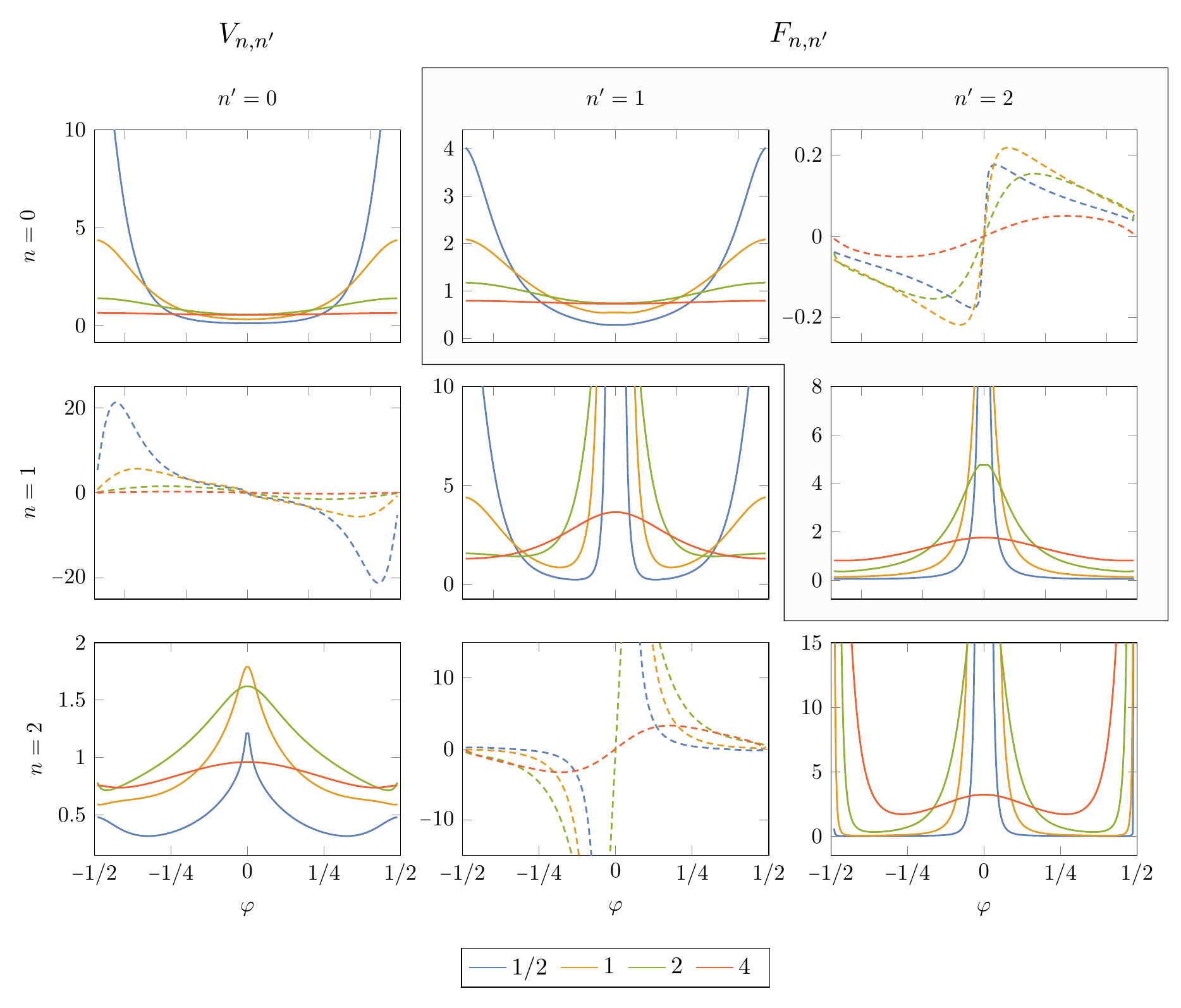}
						\caption{$V_{n, n'}$ (bottom left panels) and $F_{n, n'}$ (top right panels)   for $q = 1/2$, $1$, $2$ and $4$. Solid lines represent real-valued potentials and dashed are purely imaginary. The $V_{n, n'}$ and $F_{n, n'}$ are plotted in units of $\hbar^2/(2 f^2)$ and $\hbar/f^2$, respectively.\label{fig:newPotentials}}
			
				\end{figure}
			
				 We now turn our attention to the gapped model. The Bloch waves are the Floquet solutions of the Mathieu equation (c.f. \S\ref{sec:gappedPotential}),
				\begin{equation}
					\psi_{n, \sminus \varphi}(A) \propto \lab{me}_{2 n - 2\varphi}(\pi A, -q)
				\end{equation}
				and, as discussed above, there is a choice of $\varphi$-dependent phase that forces $F_{n, n}$ to vanish. We detail these phase conventions in (\ref{eq:mathieuBloch}) of  Appendix \ref{app:mathieu}. 
				
				While we are not able to find closed form expressions for the potentials $F_{n,n'}$ and $V_{n, n'}$, they are easily evaluated numerically. We plot a selection in Figure \ref{fig:newPotentials} for various $q$. As we argue in \S\ref{sec:gappedDynamics}, these potentials are all either purely real or purely imaginary and are related to one another by $F_{n, n'} = (-1)^{n+n'} F_{n', n}$ and $V_{n, n'} = (-1)^{n+n'}(V_{n', n} - p_\varphi F_{n', n})$.  We thus only display $F_{n, n'>n}$ and $V_{n, n' \leq n}$ to avoid unnecessary repetition. We see that neither $F_{n, n'}$ nor $V_{n, n'}$ are particularly small, and can become very large whenever the energy gap $|E_{n}(\varphi) - E_{n'}(\varphi)|$ becomes small. This makes intuitive sense---the gap fully closes when $q = 0$, and these potentials become singular in this limit.\footnote{Note that the phase conventions used in (\ref{eq:gaplessBloch2}) are not the same as those in (\ref{eq:mathieuBloch}), so that the potentials shown in Figure \ref{fig:newPotentials} (except, incidentally, those with $n = n' = 0$) do not match (\ref{eq:gaplessFnn}) and (\ref{eq:gaplessVnn}) above. It is simple, albeit tedious, to take the $q \to 0$ limit of (\ref{eq:mathieuBloch}) and repeat the computation. However, the expressions are overly long and since we will eventually analyze the gapless model using a different picture, we will just note this difference and move along. }

				In \S\ref{sec:gappedDynamics}, we will find alternative expressions for $F_{n, n'}$ and $V_{n, n'}$. We can use these to approximate the potentials in the $q \gg 1$ limit. We find\footnote{Note that the leading order, non-exponentially suppressed terms are purely real, as can be seen in Figure \ref{fig:newPotentials}. The purely imaginary potentials (i.e. $V_{1, 0}$ and $F_{0, 2}$) are always suppressed relative to these real potentials.}
				\begin{align}
					F_{n, n'} &\sim \frac{\hbar}{q^{1/4} f^2} \left(\sqrt{n'+1} \,\delta_{n, n'+1} + \sqrt{n'} \,\delta_{n, n'-1} + \mathcal{O}\big(q^{-1/2}\big)\right) +  \mathcal{O}\big(e^{-a \sqrt{q}} \cos 2 \pi \varphi\big) \label{eq:fnnApprox}\\
					V_{n, n'} &\sim \frac{\hbar^2}{2 q^{1/2} f^2}\left(\sqrt{n'(n'-1)} \,\delta_{n, n' -2} + (2 n'+1)\, \delta_{n, n'} + \sqrt{(n'+1)(n'+2)} \,\delta_{n, n'+2} + \mathcal{O}\big(q^{-1/2}\big)\right)  \nonumber \\
					&\qquad\qquad\qquad +\mathcal{O}\big(e^{-a' \sqrt{q}} \cos 2 \pi \varphi\big) \label{eq:vnnApprox}
				\end{align}
				where $a$ and $a'$ are positive constants that are not necessarily equal. For $q \gg 1$, these potentials have exponentially suppressed axion dependence. Furthermore, in this limit, $F_{n, n'}$ and $V_{n, n'}$ only connect nearest and next-to-nearest neighboring energy bands, respectively.

	\subsection{Failure of the Truncated Description} \label{sec:fail}
		
			We have derived an effective Schr\"{o}dinger equation that describes the dynamics of the axion when the gauge field is ``close'' to its vacuum state. There are, of course, friction-like couplings that allow the axion to transfer energy into the gauge field and raise it to an excited state. These will affect the evolution of the axion. But, if we truncate this system of equations and ignore these couplings, the axion simply evolves in the corrected potential $V_\lab{eff}(\varphi) + V_{0,0}(\varphi)$. Under what conditions is it reasonable to simply drop these friction-like terms?

			For compactness, let us write (\ref{eq:effEq}) as 
			\begin{equation}
				\left(i \hbar \,\partial_t - \mathcal{H}_n\right)\Phi_n = \sum_{n' \neq n} \left(F_{n, n'} p_\varphi + V_{n, n'}\right) \Phi_{n'}\,,
			\end{equation}
			with $\mathcal{H}_{n} = p_\varphi^2/(2 f^2) + E_{n}(\minus \varphi) + V_{n, n}(\varphi)$. If we assume that only $\Phi_0$ is non-vanishing initially, the $\Phi_{n \neq 0}$ are sourced entirely by $\Phi_0$ and we may safely approximate them using
			\begin{equation}
				\left|\Phi_{n}(t)\right\rangle  = \frac{1}{i \hbar} \int_{0}^{t}\!\ud t'\, e^{-i \mathcal{H}_n(t - t')/\hbar} |\Psi_0(t')\rangle\,,
			\end{equation}
			where we have defined $|\Psi_0(t) \rangle \equiv \left(F_{n, 0} p_\varphi + V_{n, 0}\right) |\Phi_0(t)\rangle.$ If the total probability in the first ``excited'' state
			\begin{equation}
				\langle \Phi_{1}(t)|\Phi_{1}(t) \rangle = \frac{1}{\hbar^2} \int_{0}^{t}\!\ud t_1 \, \ud t_2 \, \langle \Psi_{0}(t_2) | e^{-i \mathcal{H}_{1}(t_2 - t_1)/\hbar}  |\Psi_{0}(t_1)\rangle \label{eq:phi1approx}
			\end{equation}
			becomes sizable, then we should worry that the truncation no longer captures the dynamics of the full system.\footnote{This overlap is time-dependent since the Schr\"odinger equations for the $\Phi_n$ are sourced by the other axionic wavefunctions. Unitarity implies that $\sum_n \langle \Phi_n(t)|\Phi_n(t)\rangle =1$.} This will, of course, depend on the detailed dynamics of $\Phi_0$. However, from time-dependent perturbation theory, we expect schematically that\footnote{Typical time-dependent perturbation theory computes the probability of going from one state into another. A more honest analysis of this overlap involves a weighted sum over all the ways $|\Phi_0\rangle$ can transition into different eigenstates of $\mathcal{H}_1$. Because we are only interested in its typical scale, and we are summing over a complete basis of functions of $\varphi$, we approximate the left-hand ket as $\langle \Phi_0 |$.} 
			\begin{equation}
				\langle \Phi_{1}(t)|\Phi_{1}(t) \rangle \sim \frac{\left|\langle \Phi_0 | F_{1, 0} p_\varphi + V_{1, 0} |\Phi_0\rangle\right|^2}{\Delta E_{1, 0}^2}\,.
			\end{equation}
			Here, $\Delta E_{1, 0}$ is the characteristic energy difference between $\mathcal{H}_1$ and $\mathcal{H}_0$, that is
			\begin{equation}
				\Delta E_{1, 0} \sim E_{1}(\minus \varphi) + V_{1,1}(\varphi) - E_{0}(\minus \varphi) - V_{0, 0}(\varphi)\,,
			\end{equation}
			and (\ref{eq:phi1approx}) varies on a \emph{breakdown time scale} set by this energy difference, 
			\begin{equation}
					t_{\lab{break}} \sim  \frac{\hbar}{\Delta E_{1,0}}\,. \label{eq:breakdown}
			\end{equation}
			In the large $q$ limit, the energy difference $\Delta E_{1, 0}$ and the potentials $F_{1, 0}$ and $V_{1, 0}$ are roughly constant in $\varphi$, and we will approximate them as such. In general, though, they depend on the position of the axion (c.f. Figure \ref{fig:newPotentials}). If we were attempting a more accurate account of this effective description's failures, or if we were working at small $q$, we would need to keep track of this $\varphi$-dependence. We are, instead, only interested in order-of-magnitude $q \gg 1$ estimates.
			
			In the gapped model, $F_{1, 0}$ dominates over $V_{1, 0}$ and the gap is roughly $\Delta E_{1, 0} \sim 4 \sqrt{q}$. If we then compare the friction-like coupling
			\begin{equation}
					F_{0, 1} \Phi_1 \sim \frac{F_{1, 0}^2}{\Delta E_{1, 0}} p_\varphi^2 \Phi_0 \sim \frac{1}{q \,g^2 f^4} p_\varphi^2 \Phi_0
			\end{equation}
			to the effective potential
			\begin{equation}
				V_\lab{eff}(\varphi) \Phi_0 \sim \hbar^2 g^2 q^{3/4} e^{-4 \sqrt{q}}\,,
			\end{equation}
			we find a ``speed limit'' for the axion motion
			\begin{equation}
				|\dot{\varphi}|_\lab{max} \sim \hbar f^2 \left(f g\right)^2  q^{7/8} e^{-2 \sqrt{q}}\, \label{eq:speedLimit}
			\end{equation}
			by demanding that these friction terms do not dominate over the effective potential.  If the motion is dominated by the effective potential, the axion's speed is roughly $|\dot{\varphi}|_\lab{eff} \sim \hbar f^2\, (f g)\, q^{3/8} e^{-2 \sqrt{q}}$. So, we find that motion in the effective potential obeys this ``speed limit'' as long as
			\begin{equation}
				\frac{|\dot{\varphi}|_\lab{eff}\,\,}{|\dot{\varphi}|_{\lab{max}}} \sim \frac{1}{\sqrt{q}}\frac{1}{f g} \ll 1\,.
			\end{equation}

			We should mention that these are only approximate estimates, valid in the $q \gg 1 $ limit, and that a more thorough analysis of the truncated description's breakdown will depend explicitly on the axion's trajectory. We leave a detailed study of this for future work. However, we expect this estimate to hold for a variety of $V(A)$ and not just $V(A) = \Lambda(1 - \cos 2 \pi A)$, as all periodic potentials look roughly like a series of harmonic wells when considered in the tight-binding limit. Of course, these estimates fail for the gapless model, when $q = 0$, but here there is an obvious speed limit---the truncated description breaks down if the axion has enough energy to surmount the cusps (c.f. Figure \ref{fig:monodromyPotential}) at the edges of its fundamental domain.
			
			So far, we have ignored the potential mixings $V_{n, n'}$. These are more important than the friction terms $F_{n, n'}$ as they predict that the truncated description breaks down on a time scale roughly set by (\ref{eq:breakdown}), which is only polynomially suppressed by $q$, regardless of how quickly the axion moves. That is, the non-exponentially suppressed terms in (\ref{eq:ansatz}) correct the truncated description by a large amount almost immediately when compared to the non-perturbatively slow evolution in the effective potential. These terms appear because the ansatz (\ref{eq:ansatz}) is not quite right and does not account for how the dynamical axion ``backreacts'' on the gauge field.
			We might expect that they can be removed by an appropriate choice of ansatz or initial state. In the following sections, we will introduce an alternative description that allows us to treat both the axion and gauge field as a single noncompact degree of freedom. This will make the true low-energy dynamics more obvious and point toward a better choice of initial state. The axion will evolve in a corrected effective potential, still different from the one derived by equilibrium methods, while the potential mixings $V_{n, n'}$ will be exponentially suppressed. Our conclusions for the breakdown of the effective potential due to friction terms $F_{n, n'}$ will remain unchanged, even after adjusting this ansatz.

	\section{Unrolling the Compact Space}\label{sec:Unroll}
		We now show how the physics of two compact degrees of freedom---the axion $\varphi$ and the gauge field $A$---can be described by a wavefunction for a single non-compact degree of freedom. It is natural to imagine that such a description exists, as (linear) physics on a compact space is typically treated by first considering the system on a non-compact space and then performing a sum over images to enforce periodicity. This is the essence of the sum over topological sectors
		 described in \S\ref{sec:gaplessPotential}. Interestingly, the interplay between the boundary conditions (\ref{eq:boundaryConditions}) and this sum-over-images provides enough of a constraint that we only need to consider a single non-compact degree of freedom instead of two. Moreover, these images can interact when  the potential $V(A) \neq 0$.
			
		The presentation of this single degree of freedom comes in two guises, depending on which of the boundary conditions we trivialize first. In the \emph{axion frame}, the wavefunction more closely tracks the axion dynamics, while in the \emph{gauge field frame} the wavefunction more closely tracks the gauge field dynamics. Though each frame has a different Hamiltonian---and depending on the potential $V(A)$ one is usually more convenient than the other---both descriptions are fully equivalent.  After introducing these different frames and their respective Hamiltonians, we discuss how this single wavefunction encodes the state of both $\varphi$ and $A$.
		
	\subsection{Axion Frame}	
		 The wavefunction $\Psi(\varphi, A, t)$ must satisfy the boundary conditions (\ref{eq:boundaryConditions}). We may trivialize these one at a time. For instance, if we first trivialize $\Psi(\varphi, A-1, t) = \Psi(\varphi, A, t)$ by expanding in Fourier modes		
		 \begin{equation} \label{eq:axionframedef}
			\Psi(\varphi, A, t) = \sum_{\ell \in \mathbbm{Z}} \mathcal{P}_{\ell}(\varphi, t) e^{2 \pi i \ell A}\,,
		\end{equation}
		the boundary condition $e^{2 \pi i A}\Psi(\varphi-1, A, t) = \Psi(\varphi, A, t)$ implies that
			\begin{equation}
				\mathcal{P}_{\ell}(\varphi, t) = \mathcal{P}_{\ell-1}(\varphi -1, t) = \mathcal{P}_0(\varphi-\ell, t)\,, \label{eq:axionFrameWF}
			\end{equation}
			so that only a single \emph{noncompact} axionic wavefunction $\mathcal{P}(\varphi, t) \equiv \mathcal{P}_0(\varphi, t)$, $\varphi \in \mathbbm{R}$, is needed to describe the full state $\Psi(\varphi, A, t)$.\footnote{Recall that the topological term $k \times 2 \pi \hbar \varphi \dot{A}$ was necessarily quantized, $k \in \mathbbm{Z}$, and that we only consider $k = 1$ in this paper. If we instead take $k$ to be a larger integer, this theory falls into $k$ independent sectors~\cite{Seiberg:2010qd}.} 
			
			The Schr\"{o}dinger equation for $\Psi(\varphi, A, t)$ then translates into an infinite set of coupled Schr\"{o}dinger equations for the $\mathcal{P}_\ell$, 
			\begin{equation}
				\left(i \hbar\, \partial_t  - \frac{p_\varphi^2}{2 f^2} - \frac{1}{2}(2 \pi \hbar g)^2 \left(\varphi - \ell\right)^2\right) \mathcal{P}_{\ell} - \sum_{\ell' \in \mathbbm{Z}} V_{\ell'} \mathcal{P}_{\ell - \ell'} = 0\,  \label{eq:axionFrameTemp}
			\end{equation}
			where the $V_\ell$ are the Fourier coefficients of the gauge field potential,
			\begin{equation}
				V(A) = \sum_{\ell \in \mathbbm{Z}} V_\ell \,e^{2 \pi i \ell A}\,.
			\end{equation}
			Because (\ref{eq:axionFrameWF}) relates every $\mathcal{P}_\ell$ to one another, we can rewrite (\ref{eq:axionFrameTemp}) as a nonlocal Schr\"{o}dinger equation that only depends on a single axionic wavefunction $\mathcal{P}$, 
			\begin{equation}
				\left(i \hbar \,\partial_t  - \frac{p_\varphi^2}{2 f^2} - \frac{1}{2}\left(2 \pi \hbar g\right)^2 \varphi^2\right) \mathcal{P}(\varphi, t) - \sum_{\ell' \in \mathbbm{Z}} V_{\ell'} \mathcal{P}(\varphi + \ell') = 0\,. \label{eq:axionFrameSchro}
			\end{equation}
			In this noncompact description, a non-trivial gauge field potential $V(A)\neq 0$ introduces non-local interactions between the axion and itself. These non-local interactions encode the fact that the axion lives on a compact field space, i.e. that the integer-spaced points $\varphi$ and $\varphi + \ell$ are equivalent and should be able to ``talk'' to one another directly. This non-locality in the noncompact description is thus a reflection of locality in the compact description.
			
			Expectation values of the gauge-invariant operators $\exp(2 \pi i k_\lab{A} A)$ and $\exp(2 \pi i k_\varphi \varphi)$, where both $k_\lab{A}$ and $k_\varphi$ are integer, can be extracted from $\mathcal{P}(\varphi, t)$ using a non-standard Born rule,\footnote{Note that $\varphi$ is compact on the left-hand side of this equation, but noncompact on the right.}
			\begin{equation}
				\langle e^{2 \pi i k_A A + 2 \pi i k_\varphi \varphi}\rangle_\Psi = \int_{-\infty}^{\infty}\!\!\ud \varphi \, \xoverline{\mathcal{P}}(\varphi, t)\, \mathcal{P}(\varphi - k_A, t)\, e^{2\pi i k_\varphi \varphi}\,.
			\end{equation}
			Roughly, the small scale structure of $\mathcal{P}$ encodes the axion's behavior while the gauge field's behavior can be inferred from the wavefunction's large scale structure. We will make this more precise in \S\ref{sec:initialStates}, where we consider how localized Gaussian states in the compact description map into the axion frame.
			
\subsection{Gauge Field Frame}
		
\begin{figure}
			\begin{center}
			\hspace{-5pt}\includegraphics[scale=1]{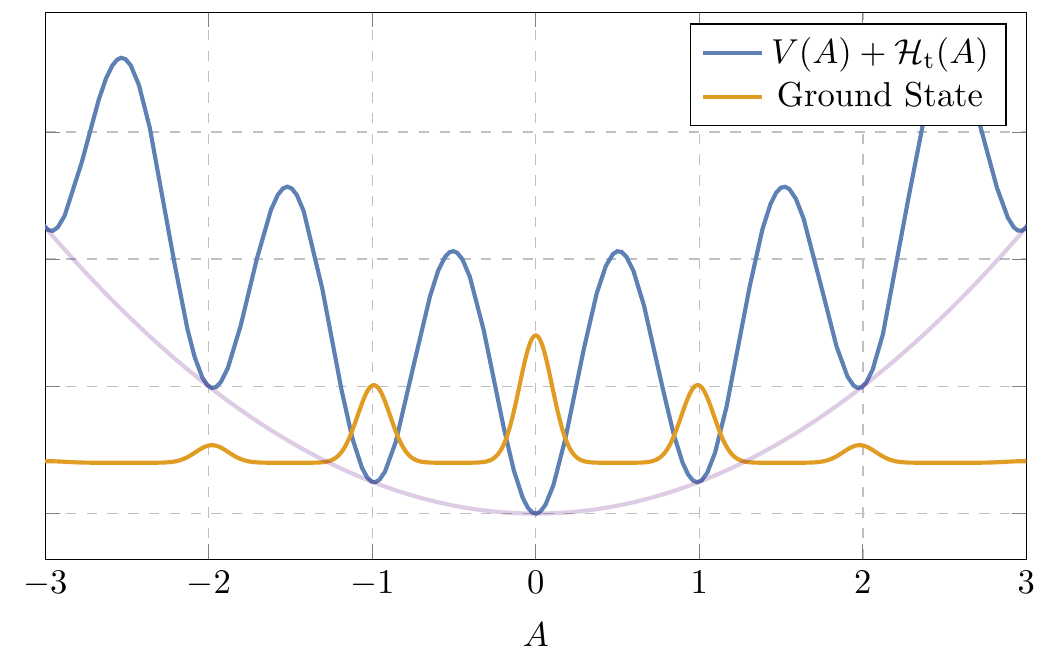}
			\caption{The gauge frame potential is a sum of a periodic potential $V(A)$ and a harmonic potential $\mathcal{H}_\lab{t}$ (light purple). For strong periodic potentials, the ground state is roughly a sum of Gaussian peaks, centered about the minima of each well, with an overall Gaussian envelope. \label{fig:gaugeFrameHam}}
			\end{center}
		\end{figure}

		An alternative description of the same system begins with trivializing the other boundary condition, $e^{2\pi i A} \Psi(\varphi-1, A, t) = \Psi(\varphi, A, t)$, by expanding in different Fourier modes
		\begin{equation}
			\Psi(\varphi, A, t) = \sum_{\ell \in \mathbbm{Z}} \mathcal{A}_\ell(A, t) e^{2 \pi i \ell \varphi + 2\pi i  A \varphi}\,.
		\end{equation}
		Again, the second boundary condition, $\Psi(\varphi, A-1, t) = \Psi(\varphi, A, t)$, forces every wavefunction $\mathcal{A}_\ell$ to be related to one another,
		\begin{equation}
			\mathcal{A}_{\ell}(A, t) = \mathcal{A}_{\ell-1}(A+1, t) = \mathcal{A}_0(A+\ell, t)\,,
		\end{equation}
		so that, again, the full state of the system $\Psi(\varphi, A, t)$ can be encoded in a single \emph{noncompact} gauge field frame wavefunction $\mathcal{A}(A, t) \equiv \mathcal{A}_0(A, t)$, with $A \in \mathbbm{R}$. In this frame, the Schr\"{o}dinger equation for $\Psi(\varphi,A,t)$ reduces to a single, uncoupled Schr\"{o}dinger equation for the wavefunction  $\mathcal{A}$,
		\begin{equation}
			i \hbar\, \partial_t \mathcal{A}(A, t) = \left(\frac{g^2 p_A^2}{2} + V(A) + \smash{\frac{1}{2}\left(\frac{ 2\pi \hbar}{f}\right)^2}\!A^2 \right) \mathcal{A}(A, t) \,, \label{eq:gfSE}
		\end{equation}
		where the gauge field frame Hamiltonian can be split into the sum of a  ``crystal'' Hamiltonian 
		\begin{equation}
		 	\mathcal{H}_\lab{c} \equiv \frac{g^2 p_A^2}{2}  + V(A) \label{eq:crystalHam}
		\end{equation}
		and a harmonic ``trap'' Hamiltonian
		\begin{equation}
			\mathcal{H}_\lab{t} \equiv \frac{1}{2}\left(\frac{2 \pi \hbar}{f}\right)^2 A^2\,. \label{eq:trapHam}
		\end{equation}
		The gauge field frame Hamiltonian---whose potential is schematically depicted in Figure \ref{fig:gaugeFrameHam}---is local even when the gauge field potential $V(A)$ is nontrivial, so it is much easier to study the system quantitatively in this frame than in the axion frame. In \S\ref{sec:gappedDynamics}, we will show how to recover the effective Schr\"{o}dinger equation (\ref{eq:effEq}) from this picture and how to improve upon it. 
		
		As in the axion frame, gauge-invariant observables are determined by a modified Born rule,
		\begin{equation}
				\left\langle e^{2 \pi i k_A A + 2 \pi i k_\varphi \varphi} \right\rangle_\Psi =  \int_{-\infty}^{\infty}\!\!\ud A \, \bar{\mathcal{A}}(A, t) \,\mathcal{A}(A - k_\varphi, t) e^{2\pi i k_A A} \label{eq:gfEV}\,.
		\end{equation}
		Now, the gauge field's behavior is determined by the small scale structure of $\mathcal{A}(A, t)$, while its large scale structure determines the axion's behavior---the opposite of the axion frame wavefunction $\mathcal{P}(\varphi, t)$. In order to make this statement more precise, we will now discuss how well-localized Gaussian states in the compact description map into the noncompact axion and gauge field frames.

		\subsection{Gaussian States} \label{sec:initialStates}
			
			A natural set of localized states to consider for noncompact $\varphi$ and $A$ are the product Gaussian states,
			\begin{align}
				\Psi_\lab{nc}(\varphi, A) \propto \mathcal{G}[\sigma_\varphi, \varphi_0, p_{\varphi, 0}](\varphi) \, \mathcal{G}[\sigma_A, A_0, p_{A, 0}](A)\,, \label{eq:compactGaussian}
			\end{align}
			where we have introduced the notation
			\begin{equation}
				\mathcal{G}[\sigma, x_0, p_0](x) \equiv \frac{1}{(\pi \sigma^2)^{1/4}}\exp\left(-\frac{1}{2 \sigma^2}(x - x_0)^2  + \frac{i p_{0}}{\hbar}(x - x_0)\right)\,.
			\end{equation}
			Clearly, $\Psi_\lab{nc}$ is a state with expected positions $(\varphi_0, A_0)$, expected momenta $(p_{\varphi, 0},  p_{A, 0})$ and whose spread is controlled by the variances $(\sigma_\varphi, \sigma_\lab{A})$. These Gaussian states are very nearly classical \cite{Hagedorn:1980et,Hagedorn:1980fp} but do not satisfy the boundary conditions (\ref{eq:boundaryConditions}), and thus do not exist in the compact Hilbert space.  
			
		Of course, we may convert any state $|\Psi\rangle_\lab{nc}$ that does not satisfy (\ref{eq:boundaryConditions}) into one that does, $|\Psi\rangle_\lab{c}$, by the method of images,
		\begin{equation}
			|\Psi\rangle_\lab{c} = \mathcal{N} \sum_{\ell_\varphi, \ell_\lab{A} \in \mathbbm{Z}}  e^{-i \ell_\varphi \pi_\varphi/\hbar - i \ell_\lab{A} \pi_A/\hbar}|\Psi\rangle_\lab{nc}\,,
		\end{equation}
		where $\mathcal{N}$ is a normalization constant.
		That is, we form a superposition of all possible translations of $|\Psi\rangle_\lab{c}$ around the torus $(\varphi, A) \sim (\varphi+1, A) \sim (\varphi, A+1)$. It is then easy to check that ``Gauss's law'' (\ref{eq:gaussLaw})
		is satisfied. Applying this sum-over-images to (\ref{eq:compactGaussian}), we find the compact space analog of a (possibly well-localized) Gaussian state
		\begin{align}
			\Psi_\lab{c}(\varphi, A) &\propto \sum_{\ell_\varphi, \ell_\lab{A} \in \mathbbm{Z}}  e^{2 \pi i \ell_\varphi A}\, \mathcal{G}[\sigma_\varphi, \varphi_0, p_{\varphi, 0}](\varphi-\ell_\varphi) \, \mathcal{G}[\sigma_A, A_0, p_{A, 0}](A-\ell_\lab{A})\,.
		\end{align}
		
		Knowing the form of $\Psi_c$, we may use it to understand how the noncompact wavefunction in either the axion or gauge field frame encodes the state of the two compact degrees of freedom. In the gauge field frame, we have
		\begin{align}
			\mathcal{A}(A) = \mathcal{N}_\lab{A} \exp\left[\minus 2 \pi^2 \sigma_\varphi^2\left(A - \frac{p_{\varphi, 0}}{2 \pi \hbar}\right)^2 - 2 \pi i \varphi_0 A \right]\sum_{\ell \in \mathbbm{Z}}\mathcal{G}[\sigma_\lab{A}, A_0, p_{A, 0}](A - \ell)\,, \label{eq:gaugeFrameIS}
		\end{align}
		and in the axion frame
		\begin{align}
			\mathcal{P}(\varphi) = \mathcal{N}_\varphi \sum_{\ell \in \mathbbm{Z}} \exp\left[-2 \pi^2 \sigma_\lab{A}^2 \left(\ell + \frac{p_{\lab{A}, 0}}{2 \pi \hbar}\right)^2 + 2 \pi i A_0 \ell \right] \mathcal{G}[\sigma_\varphi, \varphi_0, p_{\varphi, 0}](\varphi-\ell)\,, \label{eq:axionFrameIS}
		\end{align}
		where $\mathcal{N}_{\lab{A}}$ and $\mathcal{N}_\varphi$ are overall normalization constants.
		
		\begin{figure}
			\centering
			\hspace{-35pt}\includegraphics[scale=1.15, trim=0 0 0 0]{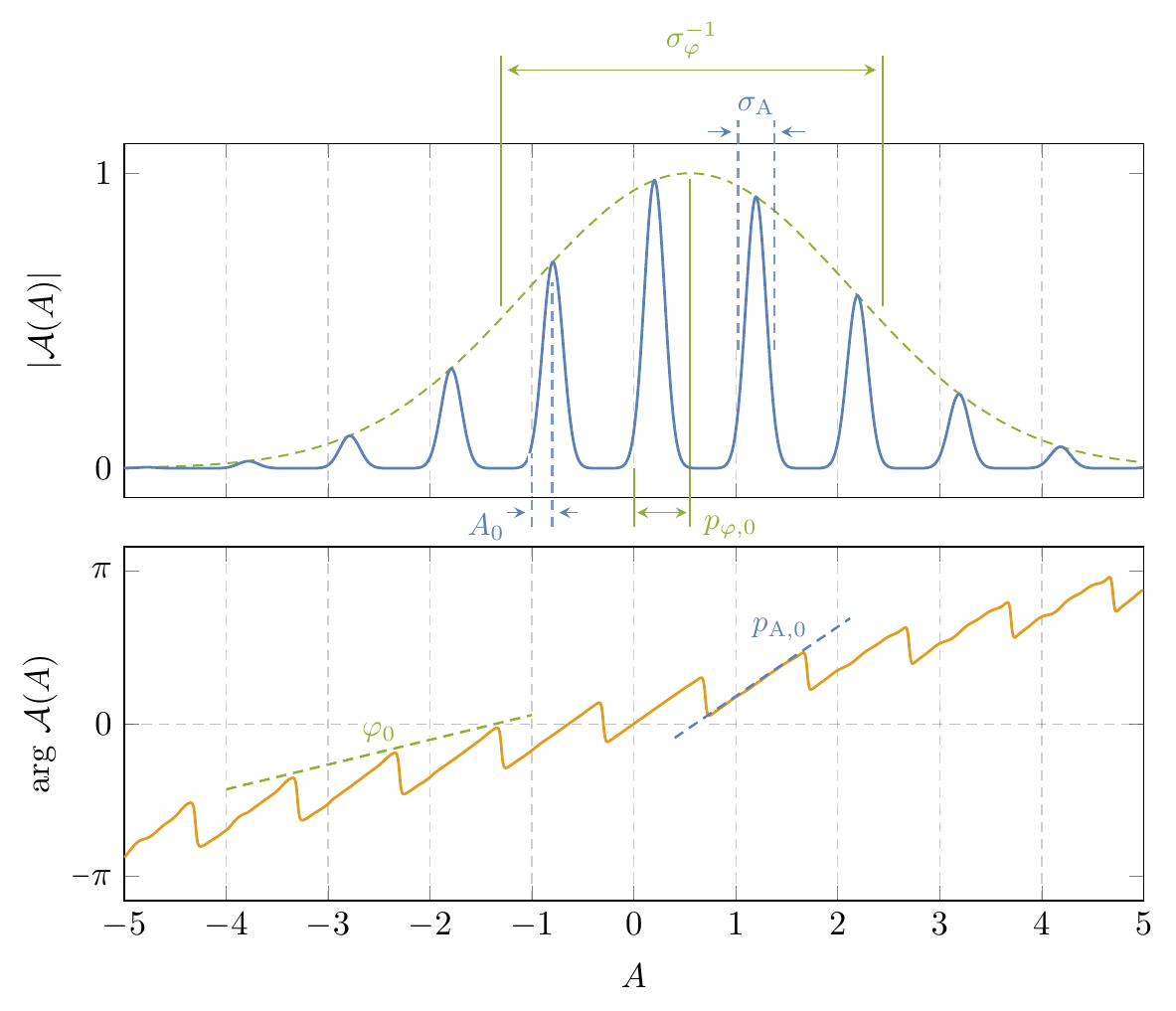}
			\caption{The magnitude and phase of a well-localized Gaussian state in the gauge field frame. As detailed in the main text, the wavefunction's small (large) scale structure determines the gauge (axion) field's state.
			}\label{fig:InState}
		\end{figure}
		
		 As illustrated in Figure \ref{fig:InState}, the compact space Gaussian wavefunction maps into a sum of evenly-spaced Gaussians  in either  frame and, though the specific structure of the wavefunction depends on the frame,  they share the same qualitative features. In the gauge field frame, the variance $\sigma_\lab{A}$ sets the width of each Gaussian peak, while the position of the gauge field $A_0$ determines the displacement of each peak from integer values of $A$. Furthermore, the momentum $p_{\lab{A}, 0}$ sets how rapidly the phase of the wavefunction varies within these peaks. These peaks are modulated by a Gaussian envelope, whose position and amplitude are determined by the momentum $p_{\varphi,0}$ and spread $\sigma_\varphi$ of the axion, respectively. Finally, the average phase difference between adjacent peaks is set by the axion's position $\varphi_0$.  Thus, the structure of the noncompact wavefunction on small scales is set by the state of the gauge field, while its behavior on large scales is determined by the axion. This is similar, but reversed, in the axion frame.
		 
		 With this dictionary in place, we can now study the dynamics of our toy model in either frame. We will begin with the gapless model in the axion frame, where we demonstrate that the axion may execute simple harmonic motion of arbitrary amplitude, and so only sees a single quadratic branch (c.f. Figure \ref{fig:monodromyPotential}) instead of the cuspy effective potential.
		 We will then use the gauge field frame to provide a more comprehensive view of axion dynamics in the gapped model.

		\section{Gapless Dynamics} \label{sec:Gaussian}
			 In \S\ref{sec:gaplessPotentials}, we found that the gapless effective Schr\"{o}dinger equation contains potentials (\ref{eq:gaplessFnn}) and (\ref{eq:gaplessVnn}) that are singular at half-integer values of $\varphi$. While the axion evolves in a simple harmonic potential within its fundamental domain, something apparently drastic happens as soon as it ventures beyond---the singular potentials excite a large number of the other wavefunctions $\Phi_{n\neq0}$.  This makes it difficult to analyze large axionic excursions. If we instead work in one of the ``unrolled'' descriptions of the previous section, the physics becomes much more transparent. Since there is no gauge field potential $V(A) = 0$, the axion frame Schr\"{o}dinger equation (\ref{eq:axionFrameSchro}) is simply that of a noncompact harmonic oscillator,
			\begin{equation}
				\left(i \hbar\,  \partial_t  - \frac{p_\varphi^2}{2 f^2} - \frac{1}{2} \left(2 \pi \hbar g\right)^2  \varphi^2 \right) \mathcal{P}(\varphi, t) = 0\,,
			\end{equation}
			with frequency $\omega = 2 \pi \hbar g/f$.
			It is then clear that the energy eigenstates in the axion frame are given by standard harmonic oscillator wavefunctions, and thus the energy eigenstates in the compact description (\ref{eq:axionframedef}) are 
			\begin{equation}
				\Psi_n(\varphi, A, t) \propto \sum_{\ell \in \mathbbm{Z}} H_{n} \big(\sqrt{2 \pi f g} (\varphi- \ell)\big) e^{-\pi f g (\varphi-\ell)^2+ 2\pi i \ell A}\,,
			\end{equation}
			with energies
			\begin{equation}
				E_n = \hbar \hspace{0.1em} \omega\!\left(\!n + \frac{1}{2}\right).
			\end{equation}
			This is not too much of a surprise. Before imposing the boundary conditions (\ref{eq:boundaryConditions}), 
			 the Hamiltonian (\ref{eq:Hamiltonian}) is that of a particle in two-dimensional flat space propagating in a magnetic field, whose Hilbert space famously arranges itself into that of an infinitely degenerate set of simple harmonic oscillators, the Landau levels. Imposing compactness then restricts this to a single harmonic oscillator.

			\begin{figure}
				\centering
				\begin{subfigure}[b]{0.35\textwidth}
					\centering
					\includegraphics[scale=1, trim=0 0 0 0]{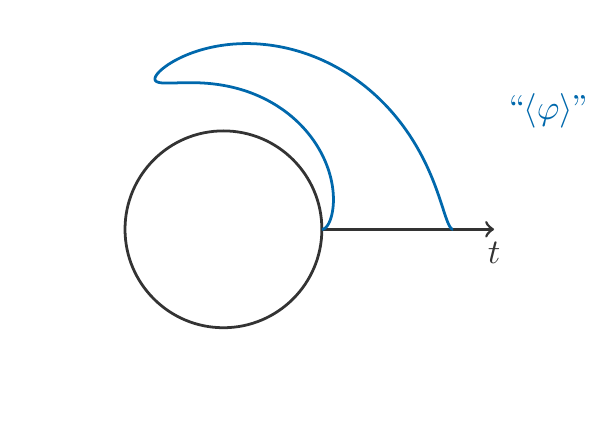}	
					\caption{
					 \label{fig:monodromyTrajectoryUnwrap}}
				\end{subfigure}
				\hspace{10pt}
				\begin{subfigure}[b]{0.35\textwidth}
					\centering
					\includegraphics[scale=1, trim=0 0 0 0]{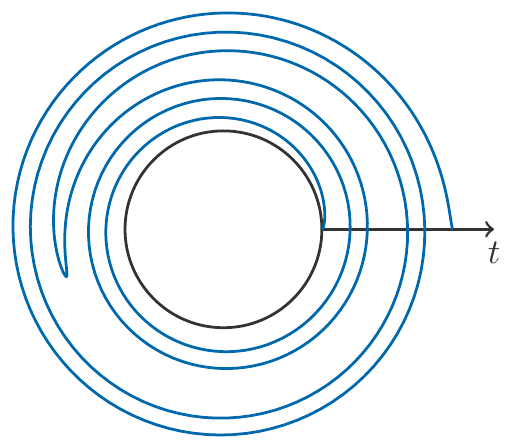}	
					\caption{
					 \label{fig:monodromyTrajectoryWrap}}
				\end{subfigure}
				\caption{For small oscillations (\ref{fig:monodromyTrajectoryUnwrap}), the axion sees the effective potential and executes simple harmonic motion. For large oscillations (\ref{fig:monodromyTrajectoryWrap}), the axion still executes harmonic motion. However, there is a monodromy. It can wrap around its field space many times before slowing to a halt and reversing course.  This motion cannot be described using an effective potential on the compact field space. \label{fig:monodromyTrajectories}}
			\end{figure}

			We are ultimately interested in the dynamics of states in which the gauge field is kept ``near'' its ground state and, since there is no potential for the gauge field, we expect that the wavefunction of these states is completely delocalized along $A$. In the previous section, we found that a Gaussian state on the compact space maps into a sum-over-Gaussians (\ref{eq:axionFrameIS}) in the axion frame. If we turn off the gauge field momentum, $p_{\lab{A}, 0} = 0$, and completely delocalize this state along $A$, $\sigma_{\lab{A}} \to \infty$, it reduces to a single Gaussian in the axion frame. Then, by taking $\sigma_{\lab{\varphi}}^{-2} = 2 \pi f g$, this becomes a simple harmonic oscillator coherent state, whose axionic expectation values time evolve very simply,
			\begin{equation}
				\left\langle \exp\left(2 \pi i \ell \varphi\right) \right\rangle = e^{-2 \pi^2 \sigma_\lab{\varphi}^2 \ell^2 }\exp\left(2 \pi i \bar{\varphi}(t)\right), 
			\end{equation}
			where
			\begin{equation}
				\bar{\varphi}(t) = \varphi_0 \cos \omega t + \frac{p_{\varphi, 0}}{\omega f^2} \sin \omega t\,.
			\end{equation}
			
\begin{figure}
			 	\centering
			 	\includegraphics[scale=1, trim= 0 0 0 0]{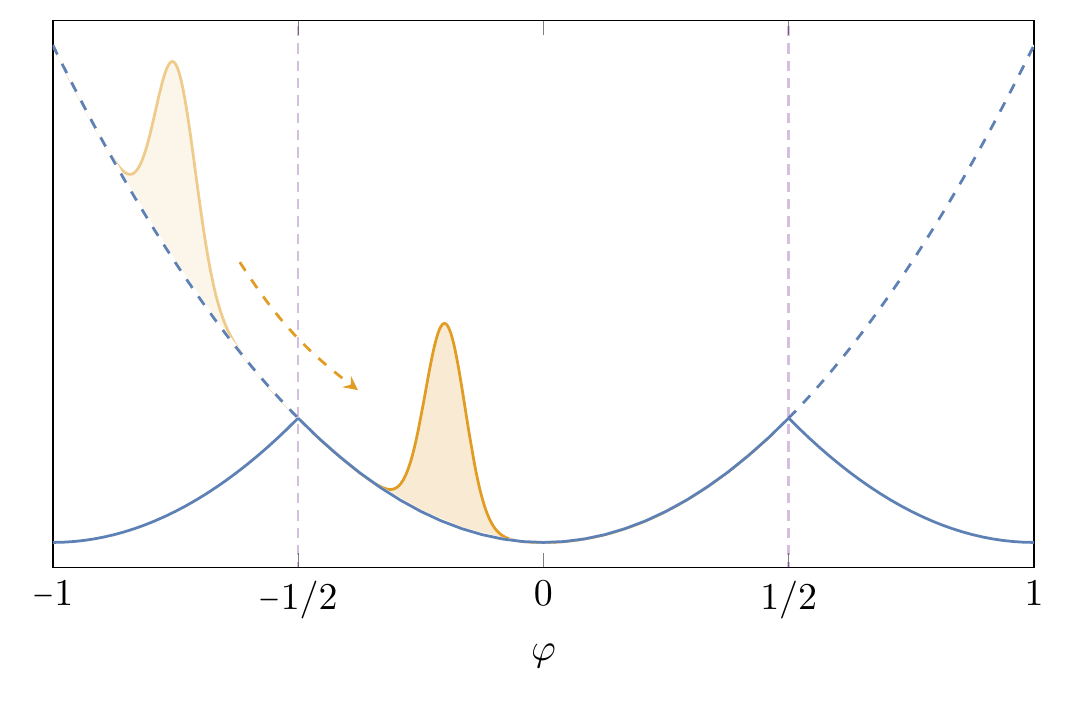}
			 	\caption{The mixings $F_{n, n'}$ and $V_{n, n'}$ force the axion onto another branch as it passes across the edge of its fundamental domain. As made obvious in the axion frame, these couplings conspire so that the axion may execute harmonic motion of arbitrary amplitude. \label{fig:axionFundamentalDomain}}
			\end{figure}

			The axion executes simple harmonic motion with amplitude $\varphi_0$.  As long as this amplitude is smaller than the fundamental domain $|\varphi_0| \leq 1/2$, the axion evolves exactly according to the effective potential, as shown in Figure \ref{fig:monodromyTrajectoryUnwrap}.  However, what is \emph{not} clear from \S\ref{sec:gaplessPotentials}'s effective description is that the axion may also execute simple harmonic motion for arbitrary amplitudes. This type of motion cannot be attributed to an effective potential on the compact field space, as the axion winds (c.f. Figure \ref{fig:monodromyTrajectoryWrap}) around its field space many times before slowing down, stopping, and turning around---there is a monodromy. As illustrated in Figure \ref{fig:axionFundamentalDomain}, the axion sees a different potential upon each revolution. If the axion instead followed the effective potential, it would wind around its fundamental domain forever, without stopping.

		This is not entirely unexpected. We may interpret this gapless model as a low-dimensional realization of the Kaloper-Sorbo mechanism \cite{Dvali:2005an, Kaloper:2008fb, Kaloper:2011jz}, i.e. a four-dimensional axion coupled to a $3$-form gauge field. Here too, the gauge field's state spontaneously breaks the axion's shift symmetry, inducing a quadratic potential. Classically, we may shift the minimum of this potential to any $\varphi$ we would like by giving the gauge field some momentum. As we might expect, this flexibility is lost at the quantum level. This can partly be seen from (\ref{eq:axionFrameIS}), where the minimum is set by the integer ``flux quantum'' (the momentum of the gauge field) $p_{\lab{A}, 0}/2 \pi \hbar$ in the $\sigma_\lab{A} \to \infty$ limit. One might wonder if this flexibility could be recovered by preparing the gauge field in a more complicated initial state, one whose momentum expectation value need not be quantized. In Appendix \ref{app:coherent}, we will rule this out by constructing a general set of coherent states. We show that, if $\varphi$ and $A$ are noncompact, there exist states that may sit still anywhere along $\varphi$'s field space. However, we then show that imposing the boundary conditions (\ref{eq:boundaryConditions}) force the axion to time evolve unless it sits at $\varphi = 0$.
			
			That the axion follows a single harmonic branch and not the effective potential, as in Figure \ref{fig:axionFundamentalDomain}, might seem like a trivial conclusion. However, we should keep in mind that this potential was generated---via a sum over topological sectors and a zero temperature $\beta \to \infty$ limit---in the same way as the gapped model (\S\ref{sec:gappedPotential}), whose effective potential we should supposedly trust for scenarios like natural inflation. The takeaway from this is that the dynamics of the axion need not follow the effective potential, which we derived by assuming the gauge field spends its entire life in a single state, and that the simplicity of the corrections in the gapless model should be attributed to the fact that it is Gaussian. Since we can continuously deform this model into the gapped model, we do not expect that these corrections simply vanish. Instead, we saw in \S\ref{sec:gappedPotentials} that their structure becomes even more non-trivial. We will now see how the axion's dynamics are described in the gauge field frame.

\section{Gapped Dynamics} \label{sec:gappedDynamics}
			While the axion frame provides a satisfying qualitative picture of how the ``localized instantons'' induced by the potential $V(A)$ affect the axion's time evolution, the non-local interactions in (\ref{eq:axionFrameSchro}) are difficult to analyze quantitatively. For non-vanishing $V(A)$, it is instead much easier to work with the gauge field frame, where our model Hamiltonian (\ref{eq:Hamiltonian}) is mapped onto that of a single particle propagating in both an infinite periodic potential (\ref{eq:crystalHam}) and a harmonic trap (\ref{eq:trapHam}). Interestingly, this system can be realized experimentally as a Bose-Einstein condensate in a optical trap \cite{Cataliotti:2001jja, Morsch:2001boa, Rey:2004uba} and has been studied both analytically \cite{Pezze:2004ibo, Hooley:2004sad, Rey:2005uca, Brand:2007eos} and numerically \cite{Ruska:2004qto, Valiente:2008qdo} by the atomic physics community. We are, however, interested in the low-energy dynamics of---from an atomic physics perspective---a type of odd-ball expectation value
			\begin{equation}
				\langle e^{2 \pi i \varphi} \rangle = \int_{-\infty}^{\infty}\!\ud A\, \bar{\mathcal{A}}(A, t) \, \mathcal{A}(A - 1, t)\,.
			\end{equation}
			In this atomic language, we are interested in the Bloch oscillations\footnote{These ``oscillations in momentum space'' were originally studied in the context of a crystal subject to a linear potential, i.e. an electric field. For a review, see \cite{Kolovsky:2004boo}.} induced by the harmonic trap, and the breakdown of our effective description is due to ``Landau-Zener tunneling.''

			 The main goals of this section are to provide both a physical motivation for the ansatz (\ref{eq:ansatz}) and an alternative derivation of the effective Schr\"{o}dinger equation (\ref{eq:effEq}), which will ultimately lead to a prescription for improving the ansatz and suppressing the potential mixings $V_{n, n'}$ described at the end of \S\ref{sec:fail}.		While our results will hold for arbitrary gauge field potential $V(A)$, we will illustrate our main points using the sinusoidal potential $V(A) = \Lambda(1 - \cos 2 \pi A)$ explored in the previous sections. It will be convenient to write the gauge field frame Hamiltonian as
			\begin{equation}
				\mathcal{H} = \frac{\pi^2 g^2 \hbar^2}{2} \left[\frac{p_\lab{A}^2}{\pi^2 \hbar^2}+ 2 q \left(1 - \cos 2 \pi A + \frac{2A^2}{q \,(f g)^2}\right)\right], \label{eq:exampleHam}
			\end{equation}
			so it is clear that the effects of the harmonic trap (and thus the effects of the axion's dynamics) are suppressed by a factor of $q (f g)^2$.
			
		\subsection{Qualitative Low Energy Dynamics}
			Before we jump into a quantitative analysis, it will be useful to first step back and understand the low-energy dynamics in (\ref{eq:exampleHam}) qualitatively. From the crystal Hamiltonian's point of view, the harmonic trap represents a singular perturbation---regardless of $q$'s size, the harmonic potential will dominate the periodic potential when {$|A| \sim  \sqrt{q}\, f g$} and force the normalizable wavefunctions to exponentially decay as $|A| \to \infty$. The Bloch waves introduced in \S\ref{sec:effPotInst} thus do not exist in (\ref{eq:exampleHam})'s Hilbert space, as these states have infinite energy. However, we expect that they remain approximate solutions in some sense, especially if we are only interested in the wavefunction's behavior near $A = 0$. We also expect that, in the limit $q \gg 1$, the low-energy wavefunctions see a potential that is roughly a sequence of evenly-spaced harmonic wells with slowly increasing minimum energy (c.f. Figure \ref{fig:gaugeFrameHam}). 
			
			\begin{figure}
				\begin{center}
					\hspace{-35pt}\includegraphics[scale=1, trim = 0 0 0 0]{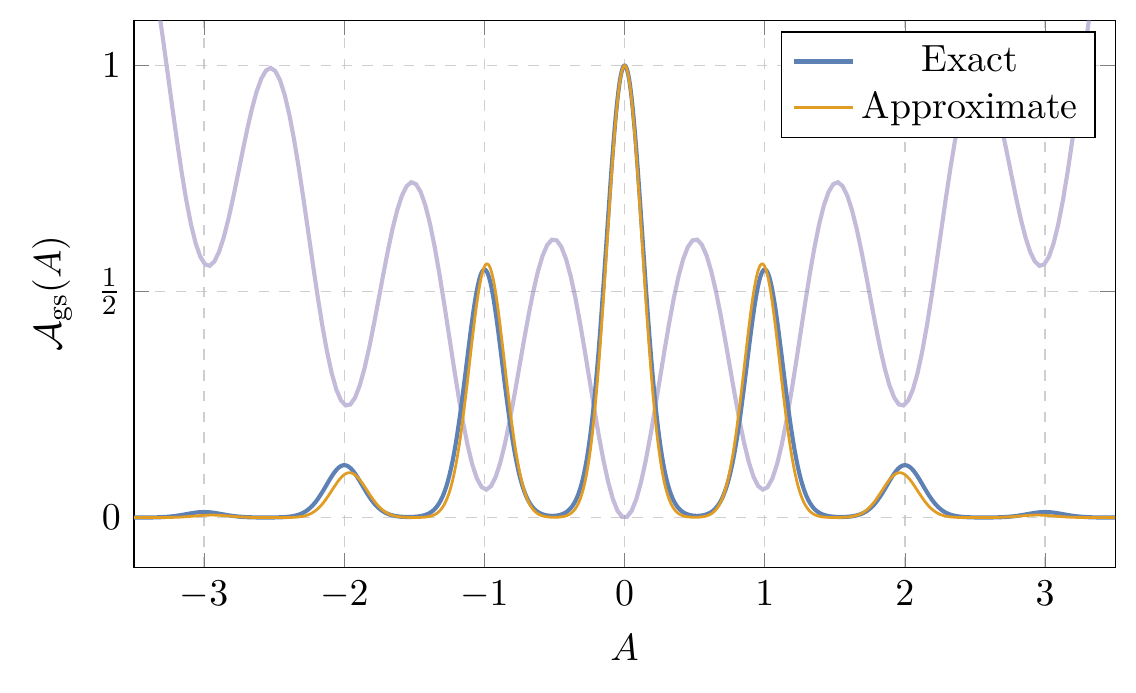}
					\caption{A comparison of the exact (i.e. found numerically) and approximate ground state wavefunction (\ref{eq:approxGS}) with $f g = 200$ and $q = 10$. We include a schematic representation of the potential (purple) as a visual guide.}
				\end{center}
			\end{figure}
			
			We thus expect that the ground state wavefunction is roughly a superposition of Gaussians centered about integer $A$ whose amplitudes exponentially decay (c.f. \ref{eq:gaugeFrameIS}),
			\begin{equation}
				\mathcal{A}_\lab{gs}(A) \sim e^{-2 \pi^2 \sigma_\varphi^2 A^2 }\sum_{k \in \mathbbm{Z}} e^{\sminus \pi^2 \sqrt{q}(A- k)^2} \label{eq:approxGS}\,.
			\end{equation}
			As seen in (\ref{eq:gaugeFrameIS}), the axion position is encoded in the phase difference between these Gaussian peaks, while its momentum is encoded in the position of the overall Gaussian envelope. It is clear that the expected position of the axion in the ground state is at $\varphi = 0$, as we can always choose the ground state wavefunction to be purely real.
			
			 We are interested in the perturbations around this ground state, which roughly divide into two classes. We should think of the perturbations that rip the gauge field from its vacuum as those that take each Gaussian in (\ref{eq:approxGS}) into another harmonic oscillator eigenstate, while perturbations that rephase the different peaks can be considered as displacements of the axion away from $\varphi = 0$. As the axion evolves in time, both the phase difference between the peaks and the center of the envelope will oscillate.
			
			From this picture, it is clear that the harmonic trap plays two roles. The first is to force the Bloch wave solutions for the crystal Hamiltonian $\mathcal{H}_\lab{c}$ to no longer be stationary states of the combined axion-gauge field system. It is the harmonic trap that forces the axion---which in the pure gauge system could be interpreted as the crystal momentum, a conserved charge---to oscillate about ${\varphi= 0}$. The second is that the harmonic trap will perturb even the local structure of the wavefunction, and we would thus expect there are additional corrections to the energy, and thus the effective potential, that disappear in the $f \to \infty$ limit. The most dominant effect is that the harmonic trap changes the concavity and location of each potential minimum near integer values of $A$. Recognizing this fact will allow us to improve the effective description (\ref{eq:effEq}).

		\subsection{Wannier Function Expansion}
				In the previous section, we argued that the low energy dynamics in the gauge field frame should map onto the rephasing of a set of almost-evenly spaced Gaussian peaks. Fortunately, the periodic Hamiltonian $\mathcal{H}_\lab{c}$ provides a complete set of functions that generalize the Gaussian states away from the $q \to \infty$ limit, known as the Wannier functions $w_{n, \ell}(A) \equiv w_{n}(A - \ell)$ and defined by
				\begin{equation}
					w_{n}(A) = \int_{\sminus 1/2}^{1/2}\!\ud \kappa \, \psi_{n, \kappa}(A)\,.
				\end{equation}
				These functions are orthonormal to one another
				\begin{equation}
					\int_{-\infty}^{\infty}\!\ud A\, w_{n, \ell} \,w_{n', \ell'}= \delta_{n, n'} \delta_{\ell, \ell'}\,,
				\end{equation}
				and with an appropriate choice of Bloch wave normalization, can be made real and  exponentially localized. It is clear from this definition that the Wannier function $w_{0}(A)$ is associated with the crystal Hamiltonian's lowest energy band.  Furthermore, it reduces to a simple Gaussian in the $q \to \infty$ limit. We collect further properties of these Wannier functions in Appendix \ref{app:wannier}.
		
				We may expand the gauge frame wavefunction\footnote{To avoid overcomplicating the following expressions, we take sums over $n, n', \dots$ to range over non-negative integers and sums over $\ell, \ell', \dots$ to range over integers.} 
				\begin{equation}
					\mathcal{A}(A, t) = \sum_{n, \ell} d_{n, \ell}(t)\, w_{n, \ell}(A) \label{eq:wannExpansion}
				\end{equation}
				and using the matrix element (\ref{eq:energyFourier}), we may rewrite the Schr\"{o}dinger equation (\ref{eq:gfSE}) as
				\begin{equation}
					i \hbar \,\dot{d}_{n, \ell} - \sum_{\ell'} E^{(\ell' \sminus \ell)}_{n} d_{n, \ell'}   - \sum_{n', \ell'} \left[\int_{-\infty}^{\infty}\!\ud A\, w_{n, \ell} \, \mathcal{H}_\lab{t}\, w_{n', \ell'} \right] d_{n', \ell'} = 0\,, \label{eq:wannSE}
				\end{equation}
				where we have introduced the Fourier coefficients of the $n$'th energy band,
				\begin{equation}
					E_{n}(\varphi) = \sum_{\ell \in \mathbbm{Z}} E^{(\ell)}_{n} e^{2 \pi i \ell \varphi}\,.
				\end{equation}
				We expect that the $n=0$ sector of (\ref{eq:wannSE}) describes the dynamics of the system when the gauge field is near its ground state.

				Now, how do we recover the effective description (\ref{eq:effEq})? Up to a normalization,  the axion expectation value (\ref{eq:gfEV}) can be expressed as
				\begin{equation}
					\langle e^{2 \pi i k \varphi} \rangle \propto \sum_{n, \ell} \bar{d}_{n, \ell} d_{n, \ell - k}\,.
				\end{equation}
				We may repackage the Wannier coefficients $d_{n, \ell}$ into periodic generating functions of a ``dummy variable'' that, with the benefit of foresight we call $\Phi_n$ and $\varphi$, respectively,
				\begin{equation}
					\Phi_n(\varphi, t) = \sum_{\ell \in \mathbbm{Z}} d_{n, \ell} \,e^{2 \pi i \ell \varphi}\,.
				\end{equation}
				At this point, the functions $\Phi_{n}(\varphi, t)$ serve as a convenient way to package the time-dependent coefficients $d_{n, \ell}(t)$. However, since  
				\begin{equation}
					\sum_{n} \int_{\sminus 1/2}^{1/2}\!\ud \varphi\, e^{2 \pi i m \varphi}\, |\Phi_{n}(\varphi, t)|^2  = \sum_{n, \ell} \bar{d}_{n, \ell} d_{n, \ell-k} \propto \langle e^{2 \pi i k \varphi} \rangle\,,
				\end{equation} 
				is the modified Born rule (\ref{eq:modBornRule}), we may identify the $\Phi_{n}$ with the axionic wavefunctions in (\ref{eq:ansatz}) and use them to reexpress the infinite set of coupled equations (\ref{eq:wannSE}) in a simpler form.
				
				By multiplying (\ref{eq:wannSE}) by $e^{2 \pi i \ell \varphi}$ and summing over $\ell$, we recover the effective Schr\"{o}dinger equation
				\begin{equation}
					i \hbar \,\partial_t \Phi_{n} = \left(\frac{p_\varphi^2}{2 f^2} + E_{n}(\minus \varphi) + V_{n, n}(\varphi)\right) \!\Phi_n + \sum_{n' \neq n} \left(F_{n, n'}(\varphi) p_\varphi + V_{n, n'}(\varphi)\right)\Phi_{n'}\, ,\label{eq:effEq2}
				\end{equation}
				but with new expressions for the potentials 
				\begin{align}
					E_{n}(\minus \varphi) &= \sum_{\ell \in \mathbbm{Z}} \left[\int_{-\infty}^{\infty}\!\ud A\, w_{n, \ell} \,\mathcal{H}_\lab{c}\, w_{n, 0}\right] e^{2 \pi i \ell \varphi}\,, \\
					F_{n, n'}(\varphi) &=  \frac{2 \pi \hbar}{f^2} \sum_{\ell \in \mathbbm{Z}} \left[\int_{-\infty}^{\infty}\!\ud A\, w_{n, \ell}\, A\, w_{n', 0} \right] e^{2 \pi i \ell \varphi} \,,\label{eq:fnnWannier}\\
					V_{n, n'}(\varphi) &=  \frac{1}{2}\left(\frac{2 \pi \hbar}{f}\right)^2\sum_{\ell \in \mathbbm{Z}} \left[\int_{-\infty}^{\infty}\!\ud A\, w_{n, \ell}\, A^2 \, w_{n', 0} \right] e^{2 \pi i \ell \varphi} \label{eq:vnnWannier}\,.
				\end{align}
				 That we recover the effective Schr\"{o}dinger equation (\ref{eq:effEq}) is no surprise as this sum over $\ell$ effectively ``rerolls'' the unrolled description, yielding simple expressions for the individual Fourier harmonics of the potentials. In fact, they can also be derived directly using the Wannier expansion (\ref{eq:wannBlochRelation}) in the expressions (\ref{eq:fnn}) and (\ref{eq:vnn}) for these potentials in terms of the Bloch waves, without a circuitous path through the gauge field frame.\footnote{As explained in Appendix \ref{app:wannier}, our choice that the $w_{n}(A)$ are real and maximally localized implies that the $F_{n, n}$ vanish and the $V_{n, n}$ are real.}

				 This rephrasing allows us to understand the structure of these potentials that would not be readily apparent from their expression in terms of the Bloch waves, nor from the dilute instanton gas approximation. For example, the reality of the Wannier functions implies that there are no non-trivial phases for higher instanton corrections to these potentials. That is, the phase of the higher Fourier harmonics is always $0$ or $\pi$.\footnote{It would be interesting to understand how the introduction of spin-orbit coupling to $\mathcal{H}_\lab{c}$, which can produce complex-valued Wannier functions, changes these conclusions.} In general, the potentials $F_{n,n'\neq n}$ and $V_{n, n'\neq n}$ are complex. However, if we assume that the gauge field potential is ``centrosymmetric,'' $V(\minus A) = V(A)$, the Wannier functions have definite parity $w_{n}(\minus A) = (-1)^n w_{n}(A)$. In this case, $F_{n, n'}$ is purely real and $V_{n, n'}$ is purely imaginary when $n+n'$ odd, while the reverse is true if $n+n'$ is even. Furthermore, $F_{n, n'} = (-1)^{n+n'+1} F_{n', n}$ and $V_{n, n'} = (-1)^{n+n'}\left(V_{n',n} - p_\varphi F_{n',n}\right)$.
				
				This picture of the dynamics also makes it clear how to improve upon the effective description. In \S\ref{sec:fail}, we found that the mixing $V_{2, 0}(\varphi)$ (c.f. \ref{eq:vnnApprox}) caused the effective description to fail quite quickly when compared to the time scale of motion in the effective potential. This failure is not due to some dynamical friction, but rather due to a description of the dynamics in the ``wrong basis." Can we alter our Wannier function expansion, and thus our choice of initial state, to suppress this term and find an effective description valid for much longer times? 
				
				In the absence of the harmonic trap, a single Wannier function $w_{n}(A)$ is an approximate stationary state of the gauge field frame Hamiltonian---its time dependence will only come from its ``leaking'' into the other potential minima of $V(A)$. However, the harmonic trap changes the concavity of the minimum, which forces $w_{n}(A)$ to oscillate in time. If we use a set of Wannier functions to take this into account, we should be able to suppress these disastrous leaking terms $V_{n,n'}$. So, instead of expanding in the Wannier functions of the crystal Hamiltonian, we may rewrite the gauge field Hamiltonian as
				\begin{equation}
					\mathcal{H} = \frac{g^2 p_\lab{A}^2}{2} + V(A) + \frac{1}{2}\left(\frac{2 \pi \hbar}{f}\right)^2 (A - \lfloor A \rceil)^2 + \frac{1}{2}\left(\frac{2 \pi \hbar}{f}\right)^2 \left[A^2 - (A - \lfloor A \rceil)^2\right]
				\end{equation}
				and expand in the Wannier functions $\tilde{w}_{n}(A)$ of the modified crystal Hamiltonian
				\begin{equation}
					\tilde{\mathcal{H}}_\lab{c} = \frac{g^2 p_\lab{A}^2}{2} + V(A) + \frac{1}{2}\left(\frac{2 \pi \hbar}{f}\right)^2 (A - \lfloor A \rceil)^2 \,.
				\end{equation}
				
				Repeating the steps that lead to (\ref{eq:effEq2}) yields effective Schr\"{o}dinger equations of the same form, yet with modified potentials,
				 \begin{align}
					\tilde{F}_{n, n'}(\varphi) &=  \frac{2 \pi \hbar}{f^2} \sum_{\ell \in \mathbbm{Z}} \left[\int_{-\infty}^{\infty}\!\ud A\, \tilde{w}_{n, \ell}\, A\, \tilde{w}_{n', 0} \right] e^{2 \pi i \ell \varphi}, \, \label{eq:fnnWannier2}\\
					\tilde{V}_{n, n'}(\varphi) &=  \frac{1}{2}\left(\frac{2 \pi \hbar}{f}\right)^2\sum_{\ell \in \mathbbm{Z}} \left[\int_{-\infty}^{\infty}\!\ud A\, \tilde{w}_{n, \ell}\, \left(A^2 - \left(A - \lfloor A\rceil\right)^2\right) \, \tilde{w}_{n', 0} \right] e^{2 \pi i \ell \varphi}\,, \label{eq:vnnWannier2}
				\end{align}
				and a corrected effective potential
				\begin{align}
						\tilde{E}_{n}(\minus \varphi) + \tilde{V}_{n, n}(\varphi) &= \sum_{\ell \in \mathbbm{Z}} \left[\int_{-\infty}^{\infty}\!\ud A\, \tilde{w}_{n, \ell} \left(\frac{g^2 p_\lab{A}^2}{2} + V(A) + \smash{\frac{1}{2}\left(\frac{2 \pi \hbar}{f}\right)^2} \!A^2\right) \tilde{w}_{n, 0}\right] e^{2 \pi i \ell \varphi}\,. \label{eq:corrEffPot}
				\end{align}
				Crucially, the integrand of (\ref{eq:vnnWannier2}) vanishes for $|A|\leq 1/2$. This, combined with the exponential suppression of the Wannier functions (\ref{eq:wannDecay}), is enough to imply that the mixings $V_{n, n'}$ are exponentially (rather than polynomially) suppressed as $q \to \infty$. Furthermore, such a change of Wannier functions cannot be used to suppress the constant part of the potential $F_{n, n'}$, and so we cannot use a change of basis to avoid the speed limit (\ref{eq:speedLimit}) we derived in \S\ref{sec:fail}. 
				
				\begin{figure}
					\centering
					\hspace{-35pt}\includegraphics{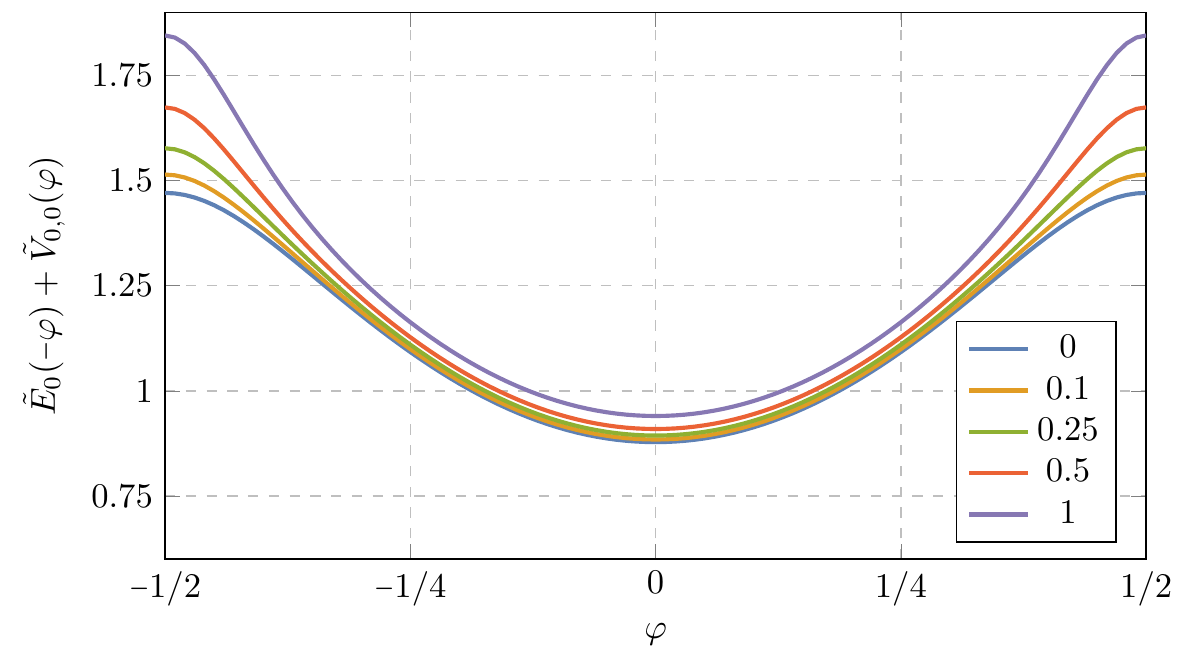}
					\caption{The corrected effective potential (\ref{eq:corrEffPot}) for $q = 0.5$ as a function of $4/(f g)^2$, in units of $(\pi \hbar g)^2/2$. The blue (bottom) curve is the uncorrected effective potential. \label{fig:corrEffPot}}
				\end{figure}		
				
				We plot a representative example of the corrected effective potential (\ref{eq:corrEffPot}) in Figure \ref{fig:corrEffPot}, for several values of $f$. As we take the decay constant $f \to \infty$ and freeze the axion in place, we recover the effective potential (the bottom blue curve) computed via standard equilibrium methods (\ref{eq:exactEnergyMathieu}). The effective potential receives corrections for finite $f$, when the axion is allowed to evolve. Near the origin $\varphi = 0$, this correction mainly raises the overall vacuum energy though there is a slight change in concavity. This has a simple interpretation. For $q \gg 1$, the crystal potential can be treated as a series of equally spaced harmonic wells. The inclusion of the harmonic trap changes the concavity of each minimum, and thus the overall vacuum energy.  As the axion approaches the edges of its fundamental domain $|\varphi| \sim 1/2$, there are additional corrections that can, depending on the dimensionless quantity $f g$, become significant.
						
		\subsection{Recap}
			Clearly, the axion's dynamics will depend on the initial state of both it and the gauge field. The challenge, then, is to find a choice of initial states which ``minimally'' excite the gauge field, so that its dynamics can be ignored and the axion can be treated as evolving in an effective potential. As we saw in \S\ref{sec:effSchrodinger}, this set of initial states is not determined by the gauge field alone---ignoring the axion's dynamics will cause large corrections to the effective description, even if the axion is sitting still. Fortunately, the gauge field frame usefully reorganizes the system's degrees of freedom, making it clear how to improve on this description. We found that the axion evolves in the \emph{corrected} effective potential (\ref{eq:corrEffPot}), shown in Figure \ref{fig:corrEffPot}, and can excite the gauge field through the couplings described by $\tilde{V}_{n, n'}$ (\ref{eq:vnnWannier2}) and $\tilde{F}_{n, n'}$ (\ref{eq:fnnWannier2}), the latter of which implies the axion speed limit (\ref{eq:speedLimit}). The relative sizes of these corrections depends on the dimensionless product $f g$ and disappear when the axion is treated as a fixed, classical parameter, as $f \to \infty$.

		\newpage
	\section{Discussion}\label{sec:Discussion}
	We have shown that the effective potential derived by equilibrium methods can fail to capture the actual semi-classical dynamics. Integrating out a degree of freedom assumes a final state and the system need not evolve into this state, so there must be corrections to the typical effective description that encode this. This paper focused specifically on non-perturbatively generated effective potentials, as they are ubiquitous in string and inflationary model building. These potentials are often used out of equilibrium, and our goal was to determine both when the equilibrium effective potential fails to capture the true dynamics of the system and how to repair the description to incorporate those ignored time-dependent effects. Said differently, our goal was to understand how to consistently integrate out non-perturbative effects when a system is driven out of equilibrium.
	
	This paper serves as a warmup to attacking this problem in quantum field theory. We focused specifically on a class of quantum mechanical toy models that, we argued, capture the relevant features of their higher-dimensional counterparts. These models feature a ``gauge field'' whose topologically non-trivial configurations generate an effective potential for a classically shift-symmetric ``axion.'' We were interested in understanding the true semi-classical dynamics of the axion when the gauge field was---in some sense---near its vacuum state and derived a set of effective Schr\"{o}dinger equations that made this precise.
	
	We showed that the axion evolves according to the effective potential but with two corrections. The first correction---a set of friction-like terms---represents the fact that the axion should be able to transfer energy to the gauge field and excite it away from its vacuum state. These appear as couplings to other wavefunctions that describe the axion propagating in the background of excited gauge field states. The second correction is to the effective potential itself, and is induced by the dynamics of the axion. The usefulness of the effective potential is usually argued on the grounds of adiabaticity---as long as the dynamics are slow enough, the integrated-out degrees of freedom should adiabatically track their vacuum and the ground state energy is a good predictor of the dynamics. This, however, is not the case. The effective potential is derived under the assumption that the axion is fixed in place. It is only accurate when the dynamics is isoaxionic (like isobaric, or isothermal) rather than solely adiabatic.
	
	We must point out that the corrections to the axion's dynamics are \emph{not} due to quantum spreading, or to quantum fluctuations of the axion. Quantum spreading of the axion is encoded in the wavefunction, and the effective Schr\"{o}dinger equation we derived is independent of our assumptions about the state of the axion. Furthermore, because we have encoded the dynamics in terms of an effective Schr\"{o}dinger equation, rather than some quantum-corrected equations of motion, we have effectively yet to do the axion's path integral.
	
	 In order to argue that this is the correct description of the \emph{low-energy} dynamics of the axion---and not some artifact from our choice of initial state---we introduced two alternative pictures of the dynamics, which we called the axion and gauge field frames. The benefit of these two frames is that they rearrange the dynamics of the two constrained degrees of freedom into a single unconstrained degree of freedom, lending a large amount of interpretative power. It was easy to see that our particular decomposition nearly captured the correct low-energy dynamics, and we showed how to repair it to yield an effective description valid for relatively long times. Furthermore, we could quantitatively estimate when this effective description breaks down.
	
	\newpage
	We have only scratched the surface of a fascinating subject, and there is much left to be done. There are many avenues for further research that would be interesting to pursue:
	
		\begin{itemize}
			\item In \S\ref{sec:fail}, we worked in the $q \gg 1$ limit where we could effectively ignore the $\varphi$-dependence of the mixing potentials $F_{n, n'}$ and $V_{n, n'}$. It would be useful to have a more precise understanding of when the effective description fails, especially depending on where the axion sits in its field space, and if it can be made more robust.
			
			\item  A simple extension of the toy model is to multiply the topological coupling by an integer $k$, keeping the periodicities (\ref{eq:periods}) fixed. This introduces a ground state degeneracy and divides the theory into discrete sectors---instead of mapping the theory of two constrained degrees of freedom onto a single one that is both noncompact and unconstrained, there will now be $k$ noncompact degrees of freedom. It would be interesting to understand the phenomenology of this model as a function of $k$. 
		
			\item We were able to realize a range of instanton behaviors by tuning the gauge field potential, smoothly interpolating between natural inflation, or instanton-like, potentials and monodromy-like potentials. There has been activity towards understanding the role of tunneling in time-dependent scenarios, particularly applied to axion monodromy \cite{Brown:2016nqt,Brown:2017wpl,Ibanez:2015fcv}. The hope is that the Weak Gravity Conjecture, when applied to the membranes that allow tunneling events, provide constraints on the maximal possible axionic excursion. However, the description of time-dependent quantum mechanical tunneling using the path integral is difficult, and requires the use of complicated Picard-Leschetz theory to understand exactly which saddles contribute~\cite{Cherman:2014sba, Andreassen:2016cff, Halliwell:2018ejl, Feldbrugge:2018gin}. The technology developed in this work, in particular the effective Schr\"{o}dinger equation and the axion frame description, may be helpful in rephrasing some of these problems and shine light on whether tunneling events imposed by the Weak Gravity Conjecture actually do restrict the maximum possible field range to be sub-Planckian.

			\item Throughout this work, we have leveraged a lot of the technology developed by condensed matter physicists to study periodic systems, i.e. crystals. Here, we only needed to consider one-dimensional crystals, but there can be qualitatively new phenomena that appear in higher-dimensional crystals. In our model, this would correspond to coupling the axion to more than one compact gauge field. It would be interesting to understand how this 
			 can qualitatively affect the Wannier functions of the system, and thus the instanton expansion of $V_\lab{eff}(\varphi)$. It would also be interesting to understand if there are systems where $V_\lab{eff}(\varphi)$ vanishes, but its correction $V_{0,0}(\varphi)$ does not.
			
			\item Does Hubble friction qualitatively change the story detailed here? It would be straightforward to couple our system to gravity in minisuperspace by introducing the scale factor as a new degree of freedom. The analysis of the resulting quantum system might teach us how the inflationary trajectories are modified by these effects.
			
			\item Coherently oscillating axions may be the cold dark matter in our universe, and it would be interesting to understand if our conclusions imply new dynamical effects in these models.
			
			\item The standard framework to analyze dynamics as a function of initial state in quantum field theory is through the Schwinger-Keldysh, or in-in, formalism. It would be very interesting to understand how to incorporate non-perturbative effects in this language.
			
		\end{itemize}
		
	While these are all interesting questions, it is most important to understand if the conclusions derived here extend to realistic quantum field theories. Since we are mainly interested in the behavior of zero modes, we do not expect our conclusions to qualitatively change by including the field's local fluctuations. If we make an analogy to quantum tunneling, it is not that the transition from quantum mechanics to quantum field theory completely disallows tunneling events. Rather, there are new considerations, like the size of vacuum bubbles, that become important.  What effects are we missing by working with these low-dimensional toy models? The reliance on non-perturbatively generated effective potentials in string and inflationary model building, particularly as a bridge between low-energy physics and quantum gravitational constraints, makes this a rather pressing question. 
	
	\subsection*{Acknowledgements}
		We would like to thank Enrico Pajer for collaboration at the early stages of this project. We also thank Nima Arkani-Hamed, Jonathan Braden, Horng Sheng Chia, Markus Dierigl, Sergei Dubovsky, Raphael Flauger, Vladimir Gritsev, Jim Halverson, Lam Hui, Cody Long, David Marsh, Liam McAllister, Miguel Montero, Alexander Polyakov, Marieke Postma, Tomislav Prokopec, Harvey Reall, Gary Shiu, Irene Valenzuela and Sebastian Zell for many helpful discussions and comments. We also thank Markus Dierigl, Miguel Montero and Enrico Pajer for helpful comments on a draft of this paper. JS thanks the Institute Henri Poincar\'{e} and Utrecht University for hospitality while parts of this work were completed. JS presented preliminary versions of this work at Cornell University, the Simons Foundation's Origins of the Universe Workshop at the IAS, Stockholm University and Utrecht University, and would like to thank the audiences for many interesting discussions and useful comments. GP is supported by the European Union's Horizon 2020 research and innovation programme under the Marie-Sk\l{}odowska Curie grant agreement number 751778. JS is supported by a Vidi grant of the Netherlands Organisation for Scientific Research (NWO) that is funded by the Dutch Ministry of Education, Culture and Science (OCW). The work of GP and JS is part of the Delta-ITP consortium.
	
	\newpage
\appendix
	
	\section{Wannier Functions}  \label{app:wannier}
		 The Bloch waves are defined as the quasi-periodic eigenstates of a periodic Hamiltonian
		\begin{equation}
			\mathcal{H}_\lab{c} \psi_{n, \kappa} = E_{n, \kappa} \psi_{n, \kappa}, \qquad\text{with}\qquad \psi_{n, \kappa}(A + 1) = e^{2 \pi i \kappa} \psi_{n, \kappa}(A)\,. \label{eq:blochWaveDef}
		\end{equation}
		This quasi-periodicity implies that we can the decompose the Bloch waves either in delocalized Fourier modes or, by a Poisson resummation, as a sum over localized \emph{Wannier functions},
		\begin{equation}
			\psi_{n, \kappa}(A) = \sum_{\ell \in \mathbbm{Z}} \psi_{n, \kappa}^\sups{(\ell)} e^{2 \pi i(\ell + \kappa) A} = \sum_{\ell' \in \mathbbm{Z}} e^{2 \pi i \kappa \ell'} w_{n}(A - \ell')\,, \label{eq:wannBlochRelation}
		\end{equation}
		where
		\begin{equation}
			w_{n}(A) = \int_{-\infty}^{\infty}\!\ud \ell\, \psi_{n, \kappa}^\sups{(\ell)} e^{2 \pi i(\ell + \kappa) A} = \int_{-\frac{1}{2}}^{\frac{1}{2}}\!\ud \kappa\, \psi_{n, \kappa}(A)\,. \label{eq:wannDef}
		\end{equation}
		We will use the notation $w_{n, \ell}(A) \equiv w_{n}(A - \ell)$ to reduce the complexity of the formulae that follow.

		 This definition does not uniquely fix the Wannier functions $w_{n, \ell}$, as the definition (\ref{eq:blochWaveDef}) leaves the overall phase of the Bloch waves $\psi_{n, \kappa}$ ambiguous. However, there always exists \cite{Kohn:1959ana, He:2001exp} a unique choice $\theta(\kappa)$ of rephasing $\psi_{n, \kappa} \to e^{i \theta(\kappa)} \psi_{n, \kappa}$ that yields a Wannier function that is real, smooth, and falls off exponentially\footnote{In the presence of spin-orbit coupling, the Wannier functions can be complex \cite{Marzari:2012max}. Similarly, if one or more bands degenerate their decay can be less than exponential. For instance, for the gapless model we study above, they decay as $|A|^{-1}$.} as
		\begin{equation}
			w_{n}(A) \sim |A|^{-3/4} e^{-h_{n} |A|} \qquad \text{as} \qquad |A| \to \infty\,, \label{eq:wannDecay}
		\end{equation}
		This choice of phase is equivalent \cite{Lensky:2014sch} to finding $w_{n, \ell}$ such that
		\begin{equation}
			\int_{-\infty}^{\infty}\!\ud A\, w_{n, \ell}\, A \, w_{n, \ell'} = 0\,.
		\end{equation}
		This is the same choice of phase that removes $F_{n, n}$ from the effective Schr\"{o}dinger equation in \S\ref{sec:effSchrodinger}.

		In the case that the periodic potential is ``centrosymmetric,'' $V(A) = V(-A)$, these Wannier functions are also (anti)symmetric. Furthermore, with the proper choice of Bloch wave normalization, 
		\begin{equation}
			\int_{-\infty}^{\infty}\!\ud A\, \bar{\psi}_{n, \kappa} \, \psi_{n', \kappa'} = \delta_{n, n'} \delta(\kappa - \kappa')\,,
		\end{equation}
		the Wannier functions at different sites are orthonormal,
		\begin{equation}
			\int_{-\infty}^{\infty}\!\ud A\, w_{n, \ell}\, w_{n', \ell'} = \delta_{n, n'} \delta_{\ell, \ell'}\,.
		\end{equation}
		
		The Fourier coefficients $E_{n}^{{(\ell)}}$ of the energy $E_{n, \kappa}$ can be represented by Wannier matrix elements of the periodic Hamiltonian $\mathcal{H}_\lab{c}$,
		\begin{align}
		\int_{-\infty}^{\infty}\!\ud A\, w_{n, \ell} \,\mathcal{H}_\lab{c}\, w_{n', \ell'} &= \delta_{n, n'} \int_{\sminus\frac{1}{2}}^{\frac{1}{2}} \!\ud \kappa \, E_{n, \kappa} e^{2 \pi i (\ell - \ell') \kappa} = \delta_{n, n'} E_{n}^{{(\ell'\sminus\ell)}}\,. \label{eq:energyFourier}
		\end{align}

		The exponential decay is determined \cite{Kohn:1959ana,He:2001exp} by the analytic structure of the energy $E_{n, \kappa}$, namely
		\begin{equation}
			\kappa_n = \begin{cases} 
				\frac{1}{2} \pm i h_{n}, & n \,\, \text{even}\\
				\pm i h_{n}, & n \,\, \text{odd}\,,
			\end{cases}
		\end{equation}
		where $E_{n, \kappa_n} = E_{n+1, \kappa_n}$. That is, the exponential decay is set by the location of the branch cut connecting the Riemann sheets of the $n$'th and $(n+1)$'th energy bands.

\section{The Mathieu Crystal} \label{app:mathieu} 
		As the literature on the Mathieu equation is full of conflicting notation,  we lay out our conventions here. Except where noted, we follow the conventions of the DLMF \cite{NIST:DLMF}. 
	
		The crystal with a perfect, cosinusoidal potential was first analyzed by Slater \cite{Slater:1952asp}. The defining equations for the Mathieu-Bloch waves are 
		\begin{equation}
			\left(\frac{1}{\pi^2}\partial_{A}^2  + \lambda_{2n + 2 \kappa}(\minus q)  + 2 q \cos 2 \pi A \right)\psi_{n, \kappa}(A) = 0
		\end{equation}
		and
		\begin{equation}
			\psi_{n, \kappa}(A + 1) = e^{2 \pi i \kappa} \psi_{n, \kappa}(A)\,.
		\end{equation}
		
		The properly normalized Bloch wave functions are 
		\begin{equation}
			\psi_{n, \kappa}(A) =  i^n
				\begin{cases}
					\lab{me}_{n+ 2\kappa}(\pi A, \minus q) & n \,\, \text{even and} \,\, \kappa \geq 0 \\
					\lab{me}_{-n-1+2\kappa}(\pi A, \minus q) & n \,\, \text{odd and} \,\, \kappa \geq 0 \\
					(-1)^n \lab{me}_{n+1+ 2\kappa}(\pi A, \minus q) & n \,\, \text{odd and} \,\, \kappa < 0 \\
					\lab{me}_{-n+2\kappa}(\pi A, \minus q) & n \,\, \text{even and} \,\, \kappa < 0
				\end{cases},\label{eq:mathieuBloch}
		\end{equation}
		where $\lab{me}_{\nu}(\pi A, -q)$ is the Floquet solution to the Mathieu equation and defined such that
		\begin{equation}
			\lab{me}_{\nu}(\pi A, 0) = e^{i \pi \nu A}\,.
		\end{equation}
		This is the choice of phase conventions that yield real Wannier functions through (\ref{eq:wannDef}).
		
		The Mathieu function has a Fourier expansion
		\begin{equation}
					\lab{me}_{\nu}(\pi A, \minus q) = \sum_{m \in \mathbbm{Z}} c_{2m}^{\nu} e^{i(2 n + \nu) \pi A}\,,  \label{eq:mathieuExpansion}
		\end{equation}
		whose coefficients satisfy the recurrence relation
			\begin{equation}
					\left( \lambda_{2n - 2 \varphi}(\minus q) - 4\left(m - \varphi\right)^2 + 2 q \right) c_{2 m} + q \left(c_{2m+2} - 2 c_{2 m} + c_{2m - 2}\right) = 0\,,
				\end{equation}
				where we write $c_{2m} = c_{2m}^{2 n \sminus 2 \varphi}$.
				For large $q$, the normalizable solutions vary slowly and this recurrence relation can be rewritten as the Schr\"{o}dinger equation for a harmonic oscillator in $x = m - \varphi$,
				\begin{equation}
					-\frac{\ud^2 c_n(x)}{\ud x^2} + \frac{1}{q} \left(4 x^2 - 2 q - \lambda_{2n - 2 \varphi}(\minus q) \right) c_n(x) = 0\,,
				\end{equation}
				with properly normalized solutions \cite{NIST:DLMF, Aunola:2003tdh}
				\begin{equation}
					c^{2n - 2 \varphi}_{2m} = (-i)^n \left(\frac{2}{\pi}\right)^{1/4} \!\!\!\frac{q^{-1/8}}{\sqrt{2^n n!}} \left(H_{n}\big(\sqrt{2}q^{-1/4} (m-\varphi)\big) e^{-(m-\varphi)^2/\sqrt{q}} + \mathcal{O}\big(q^{-1/2}\big)\right), \label{eq:mathieuFourierSolution}
				\end{equation}
				and eigenvalues $\lambda_{2n - 2 \varphi}(\minus q) \approx 2\sqrt{q}\, (2n+1) - 2 q$. From numerical experiments, this approximation is valid for large $x$ and small $n$.
				
				The Wannier functions can thus be approximated by
				\begin{equation}
					w_{n}(A) = \int_{-\infty}^{\infty}\!\ud x\, c_{n}(x) e^{2 \pi i x A} = \frac{(2 \pi)^{1/4}q^{1/8}}{\sqrt{2^n n!}}\left( H_{n}\big(\sqrt{2}\pi \,q^{1/4} A\big) e^{-\pi^2\sqrt{q} A^2} + \mathcal{O}\big(q^{-1/2}\big)\right). \label{eq:wannierApprox}
				\end{equation}
				Because the solution (\ref{eq:mathieuFourierSolution}) was valid only for large $x$, this expression for $w_{n}(A)$ is strictly only valid for small $A$. The large $A$ behavior can be determined by considering the analytic structure of $\lambda_{2n + 2\kappa}(\minus q)$ for complex $\kappa$, or by a WKB analysis. At large $A$, the lowest band Wannier function behaves as \cite{Catelani:2011raf}
				\begin{equation}
					w_{0}(A) \sim (2/\pi)^{1/4} q^{1/8} e^{-2\sqrt{q}} |A|^{-3/4} e^{- \sqrt{q} |A|}\,, \qquad \text{as} \quad A \to \pm \infty\,.
				\end{equation}

	\section{Relation to QED on a Cylinder} \label{app:qft}
		In this appendix, we describe a quantum field theory, quantum electrodynamics on a spacetime cylinder, with zero-mode sectors described by our gapless model~\cite{Hetrick:1988yg} The quantum field theories that contain the gapped model are massive deformations of this model, so we mention them briefly and refer the reader to the relevant literature, e.g. \cite{Hetrick:1995wq,Hetrick:1995yx,Hosotani:1995zg,Hosotani:1998za}.
		
		Consider $(1+1)$-dimensional QED with a single massless Dirac fermion. The Lagrangian is given by
		\begin{equation}
			\mathcal{L} = -\frac{1}{4} F_{\mu \nu} F^{\mu \nu} + \bar{\psi}\left(i \slashed{\partial} - e \slashed{A}\right) \psi \, .
		\end{equation}
		We put this theory on a spacetime cylinder of radius $L$, with boundary conditions
		\begin{equation}
			A_{\mu}(t, x+L) = A_{\mu}(t, x) \qquad \text{and} \qquad \psi(t, x+L) = -e^{2 \pi i \alpha} \psi(t, x)\,.
		\end{equation}
		We may always work in a gauge where the spatial component of the gauge field is spatially homogeneous, $A_{1}(t, x) = b(t)$. Under large gauge transformations, i.e. those that wrap non-trivially around the spatial circle,  this degree of freedom transforms as
		\begin{equation}
			b(t) \to b(t) + \frac{2 \pi}{e L}.
		\end{equation}
		This is the only degree of freedom for the gauge field, as we can eliminate $A_0$ using Gauss's law.
		
		By bosonizing the fermions, this model can be rewritten in terms of the electric charge $Q$, the axial charge $Q_5$, and a single massive bosonic degree of freedom $\bar{\phi}$ \cite{Hetrick:1988yg}, 
		\begin{equation}\label{eq:HamiltonianQEDc}
			\mathcal{H} = \frac{F^2}{2 L} + \frac{\pi}{2 L} \left(Q^2 + Q_5^2\right) + \frac{1}{2} \int_{0}^{L}\!\ud x\, \mathcal{N}_{e/\sqrt{\pi}} \left[ \bar{\Pi}^2 + (\partial_x \bar{\phi})^2 + \frac{e^2}{\pi} \bar{\phi}^2\right],
		\end{equation}
		where $\mathcal{N}_{\mu}$ represents a normal ordering in the Schr\"{o}dinger picture with respect to the mass parameter $\mu$ and $F= L \dot{b}$ is the momentum conjugate to $b$. To translate back to the original model, we use the bosonization formulae for the light-cone components of the spinor field, $\psi_\pm$, the charges $(Q, Q_5)$, and the background field strength $F$:
		\begin{align*}
			\psi_\pm = \frac{C_{\pm}}{\sqrt{L}} \exp\big(\pm i \left[q_{\pm} + 2 \pi p_{\pm}(t \pm x)/L\right]\big) :\!e^{\pm i \phi_{\pm}(t, x)}\!:\,, \qquad\qquad \\
			Q = \int_{0}^{L}\!\ud x\, \psi^\dagger \gamma^0 \psi =  -p_+ + p_- \,, \quad \text{and}\quad Q_5 =  \int_{0}^{L}\!\ud x\, \psi^\dagger \gamma_5 \psi = p_+ + p_- + \frac{e b L}{\pi}\,.
		\end{align*}
		Notice that we have separated the contribution from the zero modes $q_\pm$ of the spinors from harmonic oscillators in the bosonized chiral scalars $\phi_\pm$. 
		
		Physical states obey
		\begin{equation}
			e^{2 \pi i p_\pm} |\lab{phys} \rangle = e^{2 \pi i \alpha} | \lab{phys}\rangle\,,
		\end{equation}
		which implies that the coordinate conjugate to $p_{+} + p_{-} \equiv \bar{p}$, which we denote by $\bar{q}$, is periodic $\bar{q} \sim \bar{q} + 2\pi$. Define $p_\varphi \equiv \bar{p}/2\pi$, and $A \equiv e b L/2\pi$. The momentum conjugate to $A$ is then
		\begin{equation}
			p_\lab{A}\equiv \frac{2\pi F}{e L}\,.
		\end{equation}
		If we truncate to ``mini-superspace" with $Q=0$ and $\bar\phi=0$, we thus find that (\ref{eq:HamiltonianQEDc}) reduces to
		\begin{equation}
			\mathcal{H} = \frac{e^2 L}{8 \pi} p_\lab{A}^2 + \frac{1}{2 \pi L} \left(p_\varphi + 2 \pi A\right)^2\,.
		\end{equation}
		We may identify this as the gapless model, with
		\begin{equation}
			g^2 = \frac{e^2 L}{4 \pi} \qquad \text{and} \qquad f^2 = \pi L\,.
		\end{equation}
		
		We can ask if there are deformations of this quantum field theory that contain, in the zero-mode sector, the gapped model presented in the main text. That is possible if we introduce a mass term for the fermions in the Lagrangian. Then, the fermions induce a Coulomb potential for the gauge-field winding mode, and there is a truncation of the zero-mode sector that is the gapped model with $V(A)\propto (1-\cos(2\pi A))$ \cite{Paranjape:1993fc, Hetrick:1995wq,Hosotani:1995zg,Hosotani:1998za}.
		
	\section{Gapless Coherent States} \label{app:coherent}
		That the gapless model \S\ref{sec:gaplessPotential} is Gaussian allows us to solve it completely and, in this appendix, we will construct families of coherent states whose expectation values obey the classical equations of motion.\footnote{A smaller class of coherent states for this model has been constructed in \cite{Fremling:2013csw, Fremling:2014csw} to study the quantum hall effect on manifolds with non-trivial topology.} It will be convenient to write the gapless Hamiltonian in the generalized form
	\begin{equation}
		\hat{\mathcal{H}} = \frac{1}{2} \left(\hat{\mb{p}} - \frac{\hbar \bm{\ell} \hat{\mb{x}}}{\bar{a}^2}\right)^\top \mb{K} \left(\hat{\mb{p}} - \frac{\hbar \bm{\ell} \hat{\mb{x}}}{\bar{a}^2}\right)\,.
	\end{equation}
	where we have defined 
	\begin{equation}
		\hat{\mb{x}} = \begin{pmatrix} \hat{x}_1 \\ \hat{x}_2 \end{pmatrix} \qquad\text{and}\qquad \hat{\mb{p}} = \begin{pmatrix} \hat{p}_1 \\ \hat{p}_2 \end{pmatrix}\,,
	\end{equation}
	with generalized periodicities $x_1 \sim x_1 + a_1$ and $x_2 \sim x_2 + a_2$, the harmonic mean of these periodicities $\bar{a}^2 = a_1 a_2$,
	the kinetic matrix
	\begin{equation}
		\mb{K} = \lab{diag}\left(m_1^{-1}, m_2^{-1}\right)\,,
	\end{equation}
	and a coupling matrix
	\begin{equation}
		\bm{\ell} = \begin{pmatrix} 0 & \ell_2 \\ \ell_1 & 0 \end{pmatrix}.
	\end{equation}
	 Throughout, we will distinguish quantum operators with hats. The positions and momenta satisfy the canonical commutation relations $[\hat{\mb{x}}, \hat{\mb{p}}] = i \hbar \,\mathbbm{1}$. The benefit of this notation is that it is easily generalizable to higher dimensions, though we may recover our notation of the main text by setting $x_1 = \varphi$, $x_2 = A$, $a_1 = a_2 = 1$, $m_1 = f^2$, $m_2 = g^{-2}$, $\ell_1 = 2 \pi$, and $\ell_2 = 0$. 
	
	\subsection{General Procedure}
	The construction of coherent states for an arbitrary quadratic Hamiltonian on a noncompact space has been described in \cite{Bagrov:1990ex, Bagrov:2014cas}. The general idea is to define the coherent states as eigenfunctions of an integral of motion. That the Hamiltonian is quadratic implies that we can write this conserved quantity, the analog of the lowering operator for the harmonic oscillator, as the linear combination
	\begin{equation}
		\hat{\mb{A}} = \mb{f}\,\hat{\mb{x}}/\bar{a} + i \bar{a}\, \mb{g}\,\hat{\mb{p}}/\hbar + \bm{\varphi}\hat{\mathbbm{1}} \label{eq:loweringOp}
	\end{equation}
	where $\mb{f}$ and $\mb{g}$ are time-dependent matrices and $\bm{\varphi}$ is a time-dependent vector. We have introduced factors of $\bar{a}$ and $\hbar$ so that $\hat{\mb{A}}$, $\mb{f}$, $\mb{g}$ and $\bm{\varphi}$ are dimensionless. These time-dependent quantities are determined by requiring that $\hat{\mb{A}}$ is a conserved quantity,
	\begin{equation}
		 [i \hbar\, \partial_t - \hat{\mathcal{H}}, \hat{\mb{A}}] = 0\,, \label{eq:integOfMotion}
	\end{equation}
	and the normalization conditions
	\begin{equation}
		[\hat{\mb{A}}, \hat{\mb{A}}] = [\hat{\mb{A}}^{\!\dagger}, \hat{\mb{A}}^{\!\dagger}] = 0 \quad \text{and} \quad [\hat{\mb{A}}, \hat{\mb{A}}^{\!\dagger}] = \mathbbm{1}.
	\end{equation}
	These conditions imply that
	\begin{subequations}
		\begin{align}
			\mb{f} \mb{g}^\dagger + \mb{g} \mb{f}^\dagger = \mb{f}^\top \bar{\mb{g}} + \mb{f}^\dagger \mb{g} = \bar{\mb{f}} \mb{g}^\top + \bar{\mb{g}} \mb{f}^\top = \mb{g}^\dagger \mb{f} + \mb{g}^\top \bar{\mb{f}} = \mathbbm{1} \,,\\
			\mb{g} \mb{f}^\top - \mb{f} \mb{g}^\top = \mb{g}^\dagger \mb{g} - \mb{g}^\top \bar{\mb{g}} = \mb{f}^\dagger \mb{f} - \mb{f}^\top \bar{\mb{f}} = 0\,,
		\end{align}
	\end{subequations}
	where bars denote complex conjugation.
	
	We first define the coherent state on the non-compact space, up to an overall time-dependent normalization, as an eigenvector of this integral of motion,
	\begin{equation}
		\hat{\mb{A}} \,|\mb{z} \rangle_\lab{nc} = \mb{z} |\mb{z} \rangle_\lab{nc}\,. 
	\end{equation}
	The undetermined normalization is then fixed by requiring this state solve the Schr\"{o}dinger equation,
	\begin{equation}
		\left(i \hbar \,\partial_t - \hat{\mathcal{H}}\right) |\mb{z}\rangle_\lab{nc} = 0.
	\end{equation}
	The benefit of this formulation is that the Schr\"{o}dinger equation reduces from a partial differential equation in multiple variables to a single, ordinary differential equation in time.
	
	The complex parameter $\mb{z}$ is related to the expectation values  $\langle \mb{x} \rangle = \langle \mb{z} | \hat{\mb{x}} | \mb{z} \rangle_\lab{nc}$ and ${\langle \mb{p} \rangle = \langle \mb{z} | \hat{\mb{p}} | \mb{z} \rangle_\lab{nc}}$ by 
	\begin{equation}
		\mb{z} = \frac{\mb{f} \langle {\mb{x}}\rangle}{\bar{a}} + \frac{i \bar{a} \mb{g} \langle {\mb{p}} \rangle}{\hbar}\,,
	\end{equation}
	which may be inverted to yield
	\begin{equation}
		\langle \mb{x} \rangle = \bar{a} \left(\mb{g}^\dagger \mb{z} + \mb{g}^\top \bar{\mb{z}}\right) \qquad \text{and} \qquad \langle \mb{p} \rangle = \frac{\hbar}{i \bar{a}} \left(\mb{f}^\dagger \mb{z} - \mb{f}^\top \bar{\mb{z}}\right)\,.
	\end{equation}
	
	With the non-compact coherent state $|\mb{z}\rangle_\lab{nc}$ in hand, we may define a coherent state on the compact torus using the method of images. The torus is naturally associated with the lattice 
	\begin{equation}
		\bar{\Gamma} = \left\{ (n_1 a_1, n_2 a_2)\, | \, (n_1, n_2) \in \mathbbm{Z}_2 \right\}\,,
	\end{equation}
	and a well-defined quantum state on the torus must be invariant under all shifts in this lattice,
	\begin{equation}
		\exp\big(i\, \hat{\mb{Q}}_\lab{s} \cdot \bar{\mb{q}}/\hbar\big) |\psi \rangle = |\psi \rangle\,, \qquad \lab{for}\quad \bar{\mb{q}} \in \bar{\Gamma}\,,
	\end{equation}
	where the operator that generates these translations is 
	\begin{equation}
		\hat{\mb{Q}}_\lab{s} = \frac{\hbar}{\bar{a}^2} \bm{\ell}^\top \hat{\mb{x}} - \hat{\mb{p}}\,.
	\end{equation}
	So, we may use $|\mb{z} \rangle_\lab{nc}$ to define a \emph{compact} coherent state by summing over all translations
	\begin{equation}
		|\mb{z} \rangle_\lab{c} \equiv \mathcal{A} \,\sum_{\bar{\mb{q}} \in \bar{\Gamma}} \exp\big(i\, \hat{\mb{Q}}_\lab{s} \cdot \bar{\mb{q}}/\hbar\big) |\mb{z}\rangle_\lab{nc}, \label{eq:noncompactToCompact}
	\end{equation}
	where $\mathcal{A}$ is a normalization constant. This construction by the method of images can be shown to be equivalent to other constructions of coherent states on compact spaces \cite{Kowalski:1998hx}.
	
	\subsection{Specific Construction}
		With this general procedure in place, we now construct the coherent states for the gapless model. In what follows, we specialize to $\ell_2 = 0$. The requirement (\ref{eq:integOfMotion}) implies the equations of motion
	\begin{subequations}
		\begin{align}
			0 &= \frac{\bar{a}^2}{\hbar} \dot{\mb{g}} - i \mb{f} \mkern 1mu \mb{K}+ \mb{g} \bm{\ell}^\top \mb{K}\,, \\
			0 &= \frac{\bar{a}^2}{\hbar} \dot{\mb{f}} - \mb{f} \mkern 1mu \mb{K} \bm{\ell} - i \mb{g} \bm{\ell}^\top \mb{K} \bm{\ell}\,, \\
			0 &= \dot{\bm{\varphi}}\,.
		\end{align}
	\end{subequations}
	Without loss of generality, we can simply set $\bm{\varphi} = 0$. These equations of motion, and the initial conditions
	\begin{equation}
		\mb{f}(0) = \frac{1}{\sqrt{2}} \begin{pmatrix} \alpha_1 & 0 \\ 0 & \alpha_2 \end{pmatrix} \qquad\text{and}\qquad \mb{g}(0) = \frac{1}{\sqrt{2}} \begin{pmatrix} \alpha_1^{-1} & 0  \\ 0 & \alpha_2^{-1} \end{pmatrix}\,
	\end{equation}
	uniquely fix these coefficient matrices,
	\begin{subequations}
		\begin{align}
			\mb{f}(t) &= \frac{1}{\sqrt{2}}\begin{pmatrix} 
				\alpha_1 \cos \omega t + i \frac{\ell_1}{\alpha_1} \sqrt{\frac{m_1}{m_2}} \sin \omega t & 0  \\
				\alpha_2 \sqrt{\frac{m_1}{m_2}} \sin \omega t & \alpha_2 
			\end{pmatrix}\,, \\
			\mb{g}(t) &=  \frac{1}{\sqrt{2}} \begin{pmatrix}
				\alpha_1^{-1} \cos \omega t + i \frac{\alpha_1}{\ell_1} \sqrt{\frac{m_2}{m_1}} \sin \omega t &\,\, \minus\frac{i\alpha_1}{\ell_1} \left(1 - \cos \omega t\right) - \frac{1}{\alpha_1} \sqrt{\frac{m_1}{m_2}} \sin \omega t \\
				i \frac{\alpha_2}{\ell_1}\left(1 - \cos \omega t\right) & \frac{1}{\alpha_2} + i \frac{\alpha_2}{\ell_1} \sqrt{\frac{m_1}{m_2}} \sin \omega t
			\end{pmatrix}\,,
		\end{align}
	\end{subequations}
	where we have introduced the frequency
	\begin{equation}
		\omega = \frac{\hbar \ell_1}{\bar{a}^2\sqrt{m_1 m_2}}\,.
	\end{equation}
	With an appropriate definition of $\mb{z}$, we may then write the expectation values as
	\begin{equation}
			\langle \mb{x}(t) \rangle = \begin{pmatrix} x_{1,0} \cos \omega t + \frac{\bar{a}^2 }{\hbar \ell_1} p_{2, 0}\left(1 - \cos \omega t\right) + \frac{\bar{a}^2 m_2 }{\hbar\ell_1 m_1 }p_{1, 0}\sin \omega t \\
			x_{2, 0} - \frac{\bar{a}^2}{\hbar \ell_1 } p_{1, 0}\left(1 - \cos \omega t\right) - \frac{m_1}{m_2} \left(x_{1,0} - \frac{\bar{a}^2}{\hbar \ell_1} p_{2,0}\right) \sin \omega t \end{pmatrix} \label{eq:xexpect}
		\end{equation}
		and
		\begin{equation}
			\langle \mb{p}(t) \rangle = \begin{pmatrix}
				p_{1,0} \cos \omega t+ \frac{m_1}{m_2}\left(p_{2,0} - \frac{\hbar \ell_1}{\bar{a}^2} x_{1, 0} \right) \sin \omega t \\
				p_{2, 0}
			\end{pmatrix}\,.
		\end{equation}
		
		The defining equation for the coherent state can be written as
		\begin{equation}
		\left(\frac{i \bar{a} \mb{g}}{\hbar} \hat{\mb{p}} + \frac{\mb{f}}{\bar{a}}\left(\hat{\mb{x}} - \langle \mb{x} \rangle \right) - \frac{i \bar{a} \mb{g}}{\hbar} \langle \mb{p}\rangle \right)| \mb{z} \rangle = 0\,
	\end{equation}
	which has the obvious solution
	\begin{equation}
		\langle \mb{x} | \mb{z} \rangle_\lab{nc} = \frac{\mathcal{A}}{\sqrt{\lab{det}\, \mb{g}}} \exp\left(-\frac{1}{2 \bar{a}^2} \left(\mb{x} - \langle \mb{x} \rangle\right)^\top \mb{M} \left(\mb{x} - \langle \mb{x}\rangle\right) + \frac{i \langle \mb{p} \rangle \mb{x}}{\hbar} - \frac{i}{\hbar} \chi(t)\right)\,,
	\end{equation}
	where
	\begin{equation}
		\mb{M} = \mb{g}^{-1} \mb{f}\,.
	\end{equation}
	We will find it convenient to decompose this matrix into its real and imaginary parts, $\mb{M} = \mb{M}_\lab{r} + i \,\mb{M}_\lab{i}$. One can show that 
	this state satisfies the Schr\"{o}dinger equation if the phase $\chi(t)$ satisfies
	\begin{equation}
		\chi'(t) = \frac{m_1}{2} \langle \dot{x}_1 \rangle^2 + \frac{m_2}{2} \langle \dot{x}_2^2 \rangle + \frac{\hbar \ell_1}{\bar{a}^2} \langle x_1 \rangle \langle \dot{x}_2\rangle,
	\end{equation}
	i.e. if it is the classical action $\chi(t) = S_\lab{cl}(t)$. We may thus write the properly normalized, noncompact coherent state as
	\begin{equation}
		\langle \mb{x} | \mb{z} \rangle_\lab{nc} = \frac{1}{\sqrt{2 \pi \bar{a}^2\lab{det}\, \mb{g}}} \exp\left(-\frac{1}{2 \bar{a}^2} \left(\mb{x} - \langle \mb{x} \rangle\right)^\top \mb{g}^{-1} \mb{f} \left(\mb{x} - \langle \mb{x}\rangle\right) + \frac{i \langle \mb{p} \rangle \mb{x}}{\hbar} - \frac{i}{\hbar} S_\lab{cl}(t)\right).
	\end{equation} 
	The probability distribution covariance matrix is $\mb{M}_\lab{r}$ which at $t = 0$ is
	\begin{align}
		\mb{M}_\lab{r}(0) = \begin{pmatrix}  \alpha_1^2 & 0 \\ 0 &  \alpha_2^2 \end{pmatrix}.
	\end{align}
	The parameters $\alpha_1$ and $\alpha_2$ control the initial inverse-width of the Gaussian wavepacket in the $x_1$ and $x_2$ directions, respectively. As the coherent state evolves, this covariance matrix will evolve, and the iso-probability lines of this distribution may be described by an ellipse whose orientation and principal axes oscillate in time. It returns to its original orientation after a half-period, but with 
	\begin{equation}
		\mb{M}_\lab{r}\!\left(\frac{\pi}{\omega}\right) = \frac{\ell_1^2 }{\ell_1^2 + 4 \alpha_1^2 \alpha_2^2} \,\mb{M}_\lab{r}(0)\,.
	\end{equation}
	Generally, if the coherent state was well-localized at $t = 0$, after a half-period it will become quite delocalized. This is not so odd---the harmonic oscillator squeezed states have the same behavior, and it is only for a particular choice of initial width that the variance does not oscillate in time.
	
	With this noncompact state in hand, we may construct the compact coherent state using (\ref{eq:noncompactToCompact}), yielding
	\begin{align}
		\langle \mb{x} | \mb{z} \rangle_\lab{c} = \frac{\mathcal{A}}{\sqrt{2 \pi \bar{a}^2\,\lab{det}\, \mb{g}}} \sum_{\bar{\mb{q}} \in \bar{\Gamma}}&\exp\left(-\frac{1}{2 \bar{a}^2} \left(\mb{x}_{\bar{\mb{q}}} - \langle \mb{x} \rangle\right)^\top \mb{g}^{-1} \mb{f} \left(\mb{x}_{\bar{\mb{q}}} - \langle \mb{x}\rangle\right)\right)\nonumber \\
		&\times \exp\left( \frac{i \langle \mb{p}\rangle \cdot \mb{x}_{\bar{\mb{q}}}}{\hbar} + \frac{i \bar{\mb{q}} \cdot (\bm{\ell}^\top \mb{x}_{\bar{\mb{q}}})}{\bar{a}^2} - \frac{i}{\hbar} S_\lab{cl}(t)\right).
		\end{align}
		where we have defined $\mb{x}_{\bar{\mb{q}}} \equiv \mb{x} - \bar{\mb{q}}$. Since this is a sum over the two-dimensional lattice $\bar{\Gamma}$, it will be helpful to rewrite it in terms of a multi-dimensional Theta function, 
	\begin{equation}
			\Theta_{\bm{\alpha}, \bm{\beta}}(\mb{u}\, | \, \bm{\Omega}) = \sum_{\mb{n} \in \mathbbm{Z}^N} \exp\left(\pi i (\mb{n} + \bm{\alpha})^\top \bm{\Omega} \,(\mb{n} + \bm{\alpha}) + 2 \pi i (\mb{n} + \bm{\alpha})\cdot(\mb{u} + \bm{\beta})\right).
		\end{equation}
		where we will use the shorthand $\Theta(\mb{u}\, | \, \bm{\Omega}) = \Theta_{\bm{0}, \bm{0}}(\mb{u}\, | \, \bm{\Omega})$.
		We first introduce the basis matrix $\mb{B}$ for $\bar{\Gamma}$,
		\begin{equation}
			\mb{B} = \begin{pmatrix} a_1 & 0 \\ 0 & a_2 \end{pmatrix},
		\end{equation}
		and define the Riemann matrix, characteristic, and argument
		\begin{equation}
			\bm{\Omega} = \frac{i}{2 \pi \bar{a}^2} \mb{B}^\top \mb{g}^{-1} \mb{f}\, \mb{B}\,, \quad \bm{\alpha} = -\mb{B}^{-1}\left(\mb{x} - \langle \mb{x} \rangle\right)\, \quad \text{and} \quad \mb{u} = \frac{\mb{B}^\top}{2 \pi}\left(\frac{\bm{\ell}^\top \mb{x}}{\bar{a}^2} - \frac{\langle \mb{p} \rangle}{\hbar}\right)\,.
		\end{equation}
		The coherent state wavefunction may then be written as
		\begin{align}
			\langle \mb{x} | \mb{z} \rangle_\lab{c} = \frac{\mathcal{A}}{\sqrt{2 \pi \bar{a}^2 \, \lab{det}\, \mb{g}}} \exp\left(\frac{i \langle \mb{p} \rangle \!\cdot \!\langle\mb{x}\rangle}{\hbar}  + \frac{i \left(\mb{x} - \langle \mb{x} \rangle\right)\cdot \left(\bm{\ell}^\top \mb{x}\right)}{\bar{a}^2} - \frac{i}{\hbar} S_\lab{cl}(t)\right) \Theta_{\bm{\alpha},\mb{0}}\left(\mb{u}\, |\, \bm{\Omega}\right).
		\end{align}
		
		We may consider expectation values of the operators
		\begin{equation}
			\exp\left(2 \pi i\, \mb{q} \cdot \hat{\mb{x}}\right), \quad \text{for} \quad \mb{q} \in \Gamma,
		\end{equation}
		where $\Gamma$ is the lattice dual to $\bar{\Gamma}$, i.e. $\Gamma = \{(k_1/a_1, k_2/a_2) \, | \, (k_1, k_2) \in \mathbbm{Z}_2\}$. In the non-compact coherent state, this reads
		\begin{align}
			\langle \mb{z} | \exp\left(2 \pi i \mb{q} \cdot \hat{\mb{x}}\right) | \mb{z} \rangle_\lab{nc} = \exp\left(2 \pi i \mb{q} \cdot \langle \mb{x} \rangle - \pi^2 \bar{a}^2 \mb{q}^\top \mb{M}_\lab{r}^{-1} \mb{q}\right) \,. \label{eq:nccsev}
		\end{align}
		while in the compact space we have,
		\begin{align}
		&\frac{\langle \mb{z}| \exp\left(2 \pi i \,\mb{q} \cdot \hat{\mb{x}}\right) |\mb{z}\rangle_\lab{c}\,\,}{\langle \mb{z} | \exp\left(2 \pi i\, \mb{q} \cdot \hat{\mb{x}}\right) | \mb{z} \rangle_\lab{nc}} =  \frac{\Theta(\tilde{\mb{u}}(\mb{q}) \,|\, \tilde{\bm{\Omega}})}{\Theta (\tilde{\mb{u}}(\mb{0})\, |\, \tilde{\bm{\Omega}})}\,. \label{eq:ccsev}
		\end{align}
		Here, we have defined the notation
		\begin{subequations}
		\begin{align}
			\tilde{\bm{\Omega}} &=\frac{1}{4 \pi \bar{a}^2}\mb{B}^\top \left((\bm{\ell} + \bm{\ell}^\top) + i \left[ \mb{M}_\lab{r} + \left(\bm{\ell} + \mb{M}_\lab{i}\right)^\top \mb{M}_\lab{r}^{-1} \left(\bm{\ell} + \mb{M}_\lab{i}\right)  \right]\right)\mb{B} \nonumber \\
			&= \frac{1}{4 \pi} \begin{pmatrix} i a_1 (\alpha_1^2 + \ell_1^2/\alpha_2^2)/a_2 & \ell_1  \\ \ell_1 & i a_2 \alpha_2^2/a_1\end{pmatrix} 
			 \\
			\tilde{\mb{u}}(\mb{q}) &=  \frac{1}{2 \pi} \mb{B}^\top \left(\pi \left[\mathbbm{1} + i  \left(\bm{\ell} + \mb{M}_\lab{i}\right)^\top \mb{M}_\lab{r}^{-1}\right] \mb{q}  + \left(\frac{\bm{\ell}^\top \langle \mb{x} \rangle}{\bar{a}^2} - \frac{\langle \mb{p} \rangle}{\hbar}\right) \right).
		\end{align}
		\end{subequations}
		This compact expectation value factorizes into the non-compact result and a piece that we can attribute entirely to the compact space. We see that the only difference in the expected position between the compact and non-compact spaces will come from $\Theta(\tilde{\mb{u}}(\mb{q}) \,|\, \tilde{\bm{\Omega}})$, since the denominator $\Theta (\tilde{\mb{u}}(\mb{0})\, |\, \tilde{\bm{\Omega}})$ is time-independent.
		
		From (\ref{eq:xexpect}), the expected position of $x_1$ on the \emph{non-compact} space will not time-evolve if $p_{2, 0} = \hbar \ell_1 x_{1, 0}/\bar{a}^2$ and $p_{1, 0} = 0$. That is, $x_1$ may sit still anywhere along its field space, as long as we compensate by giving momentum to $x_{2}$. However, taking $\mb{q} = (k_1/a_1, k_2/a_2)$, we can write
		\begin{align}
			\Theta(\tilde{\mb{u}}(\mb{q}) \,|\, \tilde{\bm{\Omega}}) = \sum_{(n_1, n_2) \in \mathbbm{Z}^2} |\mathcal{A}(n_1, n_2, t)| &\times \exp\left(\pi i \left(k_1 n_1 + k_2 n_2 + \ell_1 n_1 n_2/(2\pi) \right) \right) \nonumber \\
			&\times \exp\left(i n_1 \left(\frac{\ell_1 x_{2, 0}}{a_2} - \frac{a_1 p_{1, 0}}{\hbar}\right)  - \frac{i n_2 a_2 p_{2, 0}}{\hbar} \right),
		\end{align}
		where $|\mathcal{A}(n_1, n_2, t)|$ is a real, positive, time-dependent quantity. The phase in the first line always works out to be $\pm 1$, so the only way this sum can have a non-trivial time-dependent phase is through the second line. We thus see that $x_{1}$ will oscillate whenever $p_{2, 0} \neq 0$, and so we can not have a coherent state that sits still anywhere along $x_{1}$'s field space.

		Finally, we can recover the class of coherent states mentioned in \S\ref{sec:Gaussian} by considering the class of coherent states that are maximally delocalized along the $x_2$ direction, i.e. by taking $\alpha_2 \to 0$. If we also take $\alpha_1^2 = \ell_1 \sqrt{m_1/m_2}$, we find a set of coherent states that do not spread, 
		\begin{align}
				&\langle x_1, x_2 | x_{1,0}, p_{1, 0} \rangle = \mathcal{A} \sum_{\bar{q}_1 \in a_1 \mathbbm{Z}} \exp\left(-\frac{\ell_1}{2 \bar{a}^2}\sqrt{\frac{m_1}{m_2}} \left(x_1 - \bar{q}_1 - \langle x_1 \rangle \right)^2 \right)\nonumber \\
				&\qquad \qquad \qquad	\times \exp\left(\frac{i \langle p_1 \rangle (x_1 - \bar{q}_1)}{\hbar} +  \frac{i \ell_1 \bar{q}_1 x_2}{\bar{a}^2} - \frac{i}{\hbar} S_\lab{cl}(t) \right).
			\end{align}
			Taking $a_1 = a_2 = 1$, $m_1 = f^2$, and $m_2 = g^{-2}$ recovers the variance $\sigma_\varphi^{-2} = 2\pi f g$ we derived via the axion frame.

	\newpage

\clearpage
\phantomsection
\addcontentsline{toc}{section}{References}
\bibliographystyle{utphys}
\bibliography{instooeq}

\end{document}